\pdfoutput=1
\documentclass[structabstract]{aa}  
\def \rchisq {$\chi_{\nu} ^{2}$}

\def\aj{AJ}
\def\apj{ApJ}
\def\1p5{1.5DF}

\def\apss{Astroph.Sp.Sci.}
\def\aap{A\&A}

\def\mnras{MNRAS}
\def\flux{\rm erg~s$^{-1}$~cm$^{-2}$}

\def\arcsec{$^{\prime\prime}$}
\def\msol{M$_{\odot}$}
\usepackage{txfonts}

\usepackage{latexsym,amsmath,amssymb}
\usepackage{natbib,graphicx}
\usepackage{rotating}
\bibpunct{(}{)}{;}{a}{}{,}
\def \ergsec{\hbox{erg s$^{-1}$}}
\def \hcm {\hbox {\ifmmode $ atom cm$^{-2}\else atom cm$^{-2}$\fi}}

\def \arcsec {\hbox{$^{\prime\prime}$}}

\def \chisq {$\chi ^{2}$}
\def \rchisq {$\chi_{\nu} ^{2}$}
\def\approxgt{\mathrel{\hbox{\rlap{\lower.55ex \hbox {$\sim$}}
        \kern-.3em \raise.4ex \hbox{$>$}}}}
\def\approxlt{\mathrel{\hbox{\rlap{\lower.55ex \hbox {$\sim$}}
        \kern-.3em \raise.4ex \hbox{$<$}}}}

\def\arcsec{$^{\prime\prime}$}
\def\msol{M$_{\odot}$}

\def \nh {N${\rm _H}$}
\def \ergsec{\hbox{erg s$^{-1}$}}
\def \flux{\hbox{erg s$^{-1}$ cm$^{-2}$}}
\def \hcm {\hbox {\ifmmode $ atom cm$^{-2}\else atom cm$^{-2}$\fi}}

\def \arcsec {\hbox{$^{\prime\prime}$}}

\def \chisq {$\chi ^{2}$}
\def \rchisq {$\chi_{\nu} ^{2}$}
\def\approxgt{\mathrel{\hbox{\rlap{\lower.55ex \hbox {$\sim$}}
        \kern-.3em \raise.4ex \hbox{$>$}}}}
\def\approxlt{\mathrel{\hbox{\rlap{\lower.55ex \hbox {$\sim$}}
        \kern-.3em \raise.4ex \hbox{$<$}}}}
\def \cha {$Chandra$}
\def \nh {$N{\rm _H}$}

\begin{document}

\title{Pre-outburst \cha\ Observations of the Recurrent Nova\\ T Pyxidis } 
\author{\c{S}. Balman \inst{1} \thanks{\it email:solen@astroa.physics.metu.edu.tr}}
\institute{Middle East Technical University, Dept. of Physics, Dumlup{\i}nar Bulvar{\i} Universiteler Mah. No.1, Ankara, Turkey, 06800
}
\date{}
\authorrunning{Balman \c{S}.}
\titlerunning{\cha\ Observations of T Pyx in Quiescence}


\abstract
   {}
{I study the spectral, temporal and spatial characteristics
of the quiescent X-ray emission (not in outburst) of the recurrent nova T Pyx. 
}
{I perform the spectral analysis of the X-ray data obtained using the \cha\ 
Observatory, Advanced CCD Imaging Spectrometer (ACIS-S3) detector.
I calculate the light curve of the data
and perform power spectral analysis using Fourier transform. Finally, I do  
high-resolution  imaging analysis of the
data at the sub-pixel\ level.}
{
I present a total of 98.8 ksec ($\sim$ 3$\times$30 ksec)
observation of T Pyx obtained with the ACIS-S3
detector on-board the \cha\
Observatory obtained during the quiescent phase, about 2-3 months before its outburst in April 2011.\
The total \cha\ spectrum of the source T Pyx
gives a maximum temperature kT$_{max}$$>$ 37.0 keV (2$\sigma$ lower limit) with
(0.9-1.5)$\times 10^{-13}$  \flux\ and (1.3-2.2)$\times$10$^{32}$ \ergsec\ 
(at 3.5 kpc) in the 0.1-50
keV range using a multi-temperature plasma emission model 
(i.e., CEVMKL in XSPEC). I find a
ratio of (L$_{x}$/L$_{disk}$)$\simeq$(2-7)$\times$10$^{-4}$
(assuming L$_{disk}$ $\sim$ 3$\times$10$^{35}$ erg s$^{-1}$)
indicating considerable inefficiency of emission in the boundary layer.
There is no soft X-ray blackbody emission with a 2$\sigma$ upper limit
on the blackbody temperature and luminosity
as kT$_{BB}$$<$ 25 eV and L$_{soft}$$<$ 2.0$\times$10$^{33}$ erg s$^{-1}$ in the 0.1-10.0 keV band.
All fits yield only interstellar \nh\ during quiescence.
I suggest that
T Pyx has an  optically thin boundary layer (BL) merged with an ADAF-like flow (Advection-Dominated Flow)
and/or X-ray corona in the inner disk indicating ongoing quasi-spherical 
accretion at (very) high rates during quiescent phases. Such a BL structure
may be excessively heating the WD, influencing the thermonuclear 
runaway leading to the recurrent nova events.
The orbital period of the system is detected in the power spectrum of
the \cha\  light curves. 

The central source (i.e., the binary system) emission and its spectrum is deconvolved from any
extended emission with a detailed procedure at the sub-pixel level revealing
an extended emission with S/N $\sim$6-10. The derived shape looks like an    
elliptical nebula with a semi-major axis $\sim$1.0 arcsec
and a semi-minor axis $\sim$0.5 arcsec, also indicating an elongation towards south. 
The calculated approximate count rate of the extended emission is 0.0013-0.0025 c s$^{-1}$.  
The luminosity (within errors) of the nebula is 
$\sim$(0.6-30.0)$\times$10$^{31}$ erg s$^{-1}$ (at 3.5 kpc) mostly correct towards the lower end of the range.
The nebulosity seems consistent with an interaction of the outflow/ejecta from the 1966 outburst.
}
{}


\maketitle


\section{Introduction}

A Cataclysmic Variable (CV) is a close interacting binary
system in which a white dwarf (WD) accretes material from its
late-type low mass main sequence companion.
CVs can be catagorized in two main divisions.
When the accretion occurs through an accretion
disc where magnetic field of the WD is weak or nonexistent ( $B$ $<$ 0.01
MG), such systems are referred as nonmagnetic CVs
characterized by their eruptive behavior (see the review by Warner 1995, Balman 2012).
The other class is the
magnetic CVs (MCVs) divided into two sub-classes according to the
degree of synchronization of the binary (see Mouchet et al. 2012 and references therein).

In non-magnetic CVs (T Pyx is known to belong to this class), the
material in the inner disk dissipates its
kinetic energy in order to accrete onto the slowly rotating WD creating a boundary layer (BL)
which is the transition region between the disk and the WD.
Standard accretion disk theory predicts half of the accretion luminosity to originate
from the disk in the optical and ultraviolet (UV) wavelengths and the other half to emerge from the
boundary layer as X-ray and extreme UV (EUV)/soft X-ray emission which may be summerized as
L$_{BL}$$\sim$L$_{disk}$=GM$_{WD}$$\dot M_{acc}$/2R$_{WD}$=L$_{acc}$/2 (Lynden-Bell $\&$ Pringle 1974).
During low-mass accretion rates, $\dot M_{acc}$$<$10$^{-(9-9.5)}$M$_{\odot}$\ yr$^{-1}$,
the boundary layer is optically
thin (Narayan $\&$ Popham 1993, Popham 1999) emitting mostly in the hard X-rays (kT$\sim$10$^{(7.5-8.5)}$ K).
For higher accretion rates , $\dot M_{acc}$$\ge$10$^{-9}$M$_{\odot}$\ yr$^{-1}$, the boundary layer is expected to be
opticallly thick (Popham $\&$ Narayan 1995) emitting in the soft X-rays or EUV (kT$\sim$10$^{(5-5.6)}$ K).
The transition between an optically thin and an optically thick
boundary layer does not only depend on the mass accretion rate, but
it also depends on the mass of the white dwarf (also rotation) and on the alpha
viscosity parameter.

Classical (CN) and recurrent (RN) nova outbursts occur as a result of thermonuclear runaways (i.e.,
explosive ignition of accreted material) on the surface of the
white dwarf (WD) primaries in cataclysmic variable (CV) systems ejecting material
in a range 10$^{-7}$ to 10$^{-3}$ M$_{\odot}$ with velocities from several hundred to several thousand
kilometers per second (Shara 1989; Livio 1994; Starrfield 2001; Bode $\&$ Evans 2008).
RN outbursts occur with intervals of several decades (Bode \& Evans 2008).
Nova outbursts show two main components of X-ray emission; a soft component dominating below 1 keV
originating from the hot post-outburst WD and a hard component emitting above $\sim$
1 keV as a result of accretion, wind-wind and/or blast wave interaction (Krautter 2008).
In CN and RN systems, the hard X-rays are mainly caused by the shocked plasma emission
having plasma temperatures generally in a range 0.1-10 keV with luminosities
$\le$ afew$\times$10$^{36}$ erg s$^{-1}$
{\it in the outburst stage}
(Balman, Krautter, \"Ogelman 1998, Mukai \& Ishida 2001, Orio et al. 2001; Bode et al. 2006;
Hernanz \&  Sala 2002,2007; Ness et al. 2009; Page et al. 2010;
Vaytet et al. 2011; Orlando \& Drake 2012; Nelson et al. 2012).
There has only been one resolved and detected
old CN remnant (GK Persei; Nova Per 1901)
in the X-rays studied in detail using $\sim$ 100 ksec \cha\ observation (Balman 2005; Balman \& \"Ogelman 1999).
In addition, some extension in the radial profiles of the X-ray emission was
found using the \cha\ data of the recurrent nova RS Oph, one and a half years
after the outburst possibly associated with the infrared and radio emitting regions
(Luna et al. 2009). Recently, Balman (2010) recovered extended X-ray emission
using the radial profiles of the recurrent nova T Pyx obtained from the $XMM$-$Newton$ EPIC pn data.

T Pyx had five outbursts  in
1890, 1902, 1920, 1944, and 1966 with an inter outburst time of 19$\pm$5.3 yrs
(Webbink et al. 1987). A recent quite delayed outburst occurred on April 14, 2011 (Waagan et al. 2011)
and was observed over the entire electro-magnetic spectrum including the X-rays
(e.g, Tofflemire et al. 2013, Chomiuk et al. 2014, Nelson et al. 2014).
Ground-based optical imaging of the shell of T Pyx shows
expansion velocities of about 350-500 km s$^{-1}$ (Shara et al. 1989; O'Brien \& Cohen 1998).
$Hubble\ Space\ Telescope$ (HST; 1994-2007) observations of the shell
show thousands of knots in H$\alpha$ and [NII] with expansion velocities of 500-715 km
s$^{-1}$ that have not decelerated and the main emission is within a radius of 5\arcsec-6\arcsec\
(Shara et al. 1997; Schaefer, Pagnotta \& Shara 2010). 
The spectral energy distribution (SED) is dominated by an accretion disk in the 
UV+opt+IR ranges, with a distribution (after correction for reddening) 
that is described by a power law $F_{\lambda}$=4.28$\times$10$^{-6}$ 
$\lambda$$^{-2.33}$ \flux\ ${\AA}^{-1}$, 
while the continuum in the UV 
range can also be represented by a single blackbody of T$\sim $ 34,000 K
with $\dot M$ $\sim$ (1-4)$\times$10$^{-8}$ M$_{\odot}$ yr$^{-1}$
(Gilmozzi \& Selvelli 2007; Selvelli et al. 2008).  Therefore, T Pyx is believed to be
a nonmagnetic CV accreting at high rates as expected from RN precursers with a distance
estimate of 3.50$\pm$0.35 kpc (Selvelli et al. 2008). Recently, the
ultraviolet-optical-infrared spectral energy distribution is found to be
well fitted by a power law ($f_\nu \propto \nu^1$) which suggests
that most of the T Pyx light may not be originating from a standard accretion disk, 
or any superposition of blackbodies, but rather is coming from some nonthermal source (Schaefer
et al. 2013). 

In this paper, I present analysis of three \cha\ pointings of T Pyx in the quiescent stage, few months
before the outburst in 2011. The following section is on the observations
and data reduction. Section three is on
the analysis and results of the total X-ray data. In addition, this section includes 
a detailed deconvolution of possible extended X-ray emission as seen by \cha\ and the extraction of
the central source and the plausible nebular spectrum separately.
Finally, the outcomes are discussed in the light of accretion in nonmagnetic cataclysmic variables
at high rates and also evaluation of the excess emission plausibly originating from an older nova shell.

\section{Data and Observation}
 
T Pyx was observed using the \cha\ (Weisskopf, O'dell, $\&$
van Speybroeck 1996) Advanced CCD Imaging
Spectrometer (ACIS; Garmire et al. 2000)
for a total of $\sim$ 98.8 ksec on three different pointings: 2011 January 31 (UT 16:22:36),
2011 February 2 (UT 03:05:41), and 2011 February 5 (UT 22:33:52) (PI=S. Balman). I used  S3
(the back-illuminated CCD) with the FAINT mode,
and no gratings, yielding a moderate non-dispersive energy resolution.
ACIS comprises  two CCD arrays, a 4-chip array,
ACIS-I (four front-illuminated CCDs); and a 6-chip array, ACIS-S (four
front-illuminated and two back-illuminated CCDs).

The High Resolution Mirror Assembly (HRMA) produces images with a half-power diameter (HPD)
of the point spread function (PSF) of $<$ 0$^{\prime\prime}$.5 .
ACIS has an unprecedented angular resolution of 0$^{\prime\prime}$.49 per pixel.
The encircled power radii of ACIS at
50$\%$ and 80$\%$  are 0$^{\prime\prime}$.418 and 0$^{\prime\prime}$.685, respectively.
The 0$^{\prime\prime}$.418 resolution of the ACIS PSF core-radius is
exploit to recover any X-ray nebulosity that is extended.

The pipeline-processed data (aspect-corrected, bias-subtracted, graded and
gain-calibrated event lists) are used for the analysis, and
$acis$-$process$-$events$ thread is used to double check/reprocess the
level 1 data using the necessary calibration files with the aid of CIAO 4.3
and a suitable CALDB 4.4.2 when necessary. For further analysis, HEASOFT version 6.9-6.13
is utilized. In order to double check, the archived data of T Pyx in
February 2012 is also analyzed using CIAO 4.4 and  CALDB 4.4.8.
Some of the preliminary results regarding any extended emission
can be found in Balman et al. (2012).

\section{Analysis and Results}

\subsection{The \cha\ Spectrum of the Total Observation}

To obtain a total spectrum from the three data sets of T Pyx, first, I derived
the source+background spectra and the background spectra (using CIAO task $specextract$)
for the individual observations using a
circular photon extraction radius of 5\arcsec (extraction area shown in Figure 3a--the large circles). 
Next, I combined the spectra using the
task $combine$-$spectra$ which also calculates combined response matrix and ancillary response
files.  In order to understand the origin/s of this total X-ray spectrum, I fitted the spectrum
with suitable plasma emission models in XSPEC. The basic results are given in Table 1 and some of the
fitted spectra are displayed in Figure 1.

\begin{table}[ht]

\label{1}
\caption{ Spectral Parameters of the Total (combined) Spectrum of T Pyx (0.2-9.0 keV);
ranges correspond to 90$\%$ confidence level errors ($\Delta$\chisq = 2.71 -- single parameter)
$\chi^2_{\nu}$\ values of the fits are $\le$ 1.0.}
\begin{center}
\begin{tabular}{c|c|c} \hline\hline
\multicolumn{1}{c}{  } &
\multicolumn{1}{c}{\ CEVMKL$^{\S{1}}$} &
\multicolumn{1}{c}{\ MEKAL$^{\S{2}}$} \\
\hline

 \nh$_1$ ($\times 10^{22}$ cm$^{-2}$) &  0.1$^{+0.10}_{-0.05}$ &
0.3$^{+0.2}_{-0.2}$  \\

${\alpha}$  &  0.7$^{+0.3}_{-0.2}$ &
N/A \\

kT$_{1}$ (keV) &  N/A &
0.25$^{+0.13}_{-0.05}$ \\

kT$_{max}$ (keV) &  37.0$<$ &
N/A \\

K$_{cevmkl}$  ($\times 10^{-5}$ cm$^{-5}$) & 4.4$^{+0.7}_{-0.4}$ & N/A  \\

K$_{mekal}$$^{\S{3}}$\  ($\times 10^{-6}$ cm$^{-5}$) &
N/A & 4.0$^{+2.4}_{-2.0}$  \\

\hline
\hline

kT$_{2}$ (keV) & N/A  &
 22$<$  \\

K$_{mekal}$$^{\S{3}}$\  ($\times 10^{-5}$ cm$^{-5}$) &
N/A & 3.7$^{+0.4}_{-1.7}$  \\

\hline
\end{tabular}
\end{center}
{\bf Notes.}
{\bf \S{1}} {The  model is ($tbabs$*CEVMKL);
$tbabs$--Wilms et al. 2000; CEVMKL is a multi-temperature plasma emission model built from the mekal code.
Emission measures follow a power-law in temperature (dEM = (T/Tmax)$^{\alpha-1}$ dT/Tmax).}\ 
{\bf \S{2}} {The  model is ($tbabs$*(MEKAL+MEKAL));
MEKAL--Mewe et al. 1986.}\ 
{\bf \S{3}} {The normalization constant of the MEKAL plasma emission models is
K=(10$^{-14}$/4$\pi$D$^2$)$\times$EM where EM (Emission Measure) =$\int n_e\ n_H\ dV$
(integration is over the emitting volume V). The lower limits on temperature parameters are at 
2$\sigma$ significance level.}
\end{table}

The X-ray spectra of nonmagnetic CVs is found to show
a temperature distribution of hot optically thin 
plasma emission as the accreting gas settles on the WD through the boundary layer
revealing several emission lines detected
ranging from H- and He-like elements to Fe L-shell lines
(Mukai et al. 2003,
Baskill et al. 2005, Pandel et al. 2005, Guver et al. 2006,
Okada et al. 2008, Balman et al. 2011, Balman 2014).
These spectra are well modeled with an isobaric cooling flow that is a multi-temperature
distribution of plasma with differential emission measure assuming power-law distribution
of temperatures ($dEM=(T/T_{max})^{\alpha -1} 
dT/T_{max}$; e.g. MKCFLOW or CEVMKL within XSPEC software).

The combined
spectrum of T Pyx can not be fitted with a single-temperature component plasma emission 
model in collisional
equilibrium or a blackbody model, with \rchisq\ values much larger than 2.
Figure 1 displays the total spectrum of the source which portrays a smooth continuum-like
characteristic. This is due to the low count rate of the source which yields only about 510
net source counts in the total $\sim$100 ksec \cha\ observation. For any given plasma model with 
near-solar composition and moderate emissivity from lines convolved with moderate spectral resolution
of the ACIS-S detector (with no gratings in use), these features may be smeared out
and superimposed over the continuum. Therefore, non-detection of particular emission lines are
expected.

A (cooling flow-type)
multi-temperature distribution plasma model, (with a power-law distribution of temperatures)
CEVMKL in $XSPEC$, has been used to model the spectrum of T Pyx in accordance with
the general concensus of the X-ray spectra of nonmagnetic CVs as mentioned in the previous paragraphs.
The fit result yield
N$_{\rm H}$=0.10$^{+0.10}_{-0.05}$$\times 10^{22}$ cm$^{-2}$, kT$_{max}$= 100$^{<}_{-53}$ keV,
$\alpha$=0.7$^{+0.3}_{-0.2}$ (power law index of the temperature distribution) and the
normalization is 4.4$^{+0.7}_{-0.4}$$\times 10^{-5}$ (see Table 1 and Figure 1a).
The unabsorbed X-ray flux is (5.0-8.0)$\times 10^{-14}$  \flux\ in the 0.2-9.0
keV range which translates to a luminosity of (0.8-1.2)$\times 10^{32}$ \ergsec\ 
(at 3.5 kpc source distance).
The flux and luminosity are (0.9-1.5)$\times 10^{-13}$  \flux\ and (1.2-2.2)$\times 10^{32}$ \ergsec\ 
in a wider energy band of 0.1-50.0 keV.
The maximum X-ray temperature has a very high
unconstrained best-fit value of kT$_{max}$ $>$ 47 keV with a 2$\sigma$ lower limit of 37 keV.
The power law index $\alpha$ of the X-ray temperatures diverges somewhat from 1.0 (an index
$\alpha$ of 1.0 is expected from an isobaric cooling flow-type plasma). 
The N$_{\rm H}$ value derived from the fit is consistent (at 95\% confidence level)
with the interstellar value derived for T Pyx using $nhtot$,
0.28$\times 10^{22}$ cm$^{-2}$ from the database of
Willingale et al. (2013) (http://www.swift.ac.uk/analysis/nhtot/index.php) who use
the atomic hydrogen column density, N(HI),
and the dust extinction, E(B-V), describing the variation of the molecular
hydrogen column density, N(H2), of our Galaxy, over the sky using 21 cm radio emission maps and the
$Swift$ GRB data. The measured value of E(B-V)=0.5-0.25 (higher limit from
Shore et al. 2011:during outburst, lower limit from Schafer et al. 2013:during quiescence) yields
a range of N$_{\rm H}$=(3-1.5)$\times 10^{21}$ cm$^{-2}$ consistent with the $nhtot$ result and the
derived spectral parameter N$_{\rm H}$ in Table 1.

The normalization of the CEVMKL model is similar to MEKAL assuming an average emission measure (EM)
(see also Table 1).
Taking a distance of 3.5 kpc and that $<EM>$=$<n_e>^2 V$ (V=emitting volume), an electron density can be
approximated. For an emitting volume of (3$\times$10$^{9}$cm)$^3$ for simplicity, the best-fit value of 
the normalization gives an average electron density of 10$^{13}$\ cm$^{-3}$. This density yields 
equipartition of temperature between electrons and ions (CIE) in about 10 seconds assuming 37 keV
temperature and if the temperature is lower the equipartition is faster 
(equipartition timescale from Fransson et al. 1996).
During the fitting process a solar plasma composition has been assumed. 
When the abundances are set free in the CEVMKL fits, some over abundances
in  the elements are seen but not detected with any significance and the \rchisq\ of the fit is not 
altered. I have also checked to the general metal abundances 
using the CEMEKL model (CEVMKL without individual abundances). Such a fit yields almost the same 
spectral parameters as in the CEVMKL model fit with the additional parameter range of metal 
abundance of 0.4-2.3 (90\% confidence level) for the quiescent spectrum of T Pyx.  
In general, I note that the X-ray temperature 
is very high and the power law index of the temperature distribution
slightly diverge from an isobaric cooling flow
for an optically thin hard X-ray emitting boundary layer. At the
accretion rates of T Pyx $\dot{\rm M}$\ $>$10$^{-8}$ M$_{\odot}$ yr$^{-1}$, the boundary layer
is expected to be in the optically thick regime (see Popham \& Narayan 1995,
Popham 1999).

A double plasma emission model can be fitted with \rchisq\ around 1.0 yielding spectral parameters;
an N$_{\rm H}$ of  0.3$^{+0.16}_{-0.22}$$\times 10^{22}$ cm$^{-2}$, a kT1 of 0.25$^{+0.13}_{0.05}$,
a normalization 4.0$^{+2.4}_{-2.0}$$\times 10^{-6}$,
and a kT2 of 79.0$^{<}_{-57.0}$ keV with a normalization of
3.7$^{+0.4}_{-1.0}$$\times 10^{-5}$ (errors are given at 90\% confidence level).
The integrated unabsorbed X-ray flux is (5.0-9.0)$\times 10^{-14}$  \flux\ in the 0.2-9.0
keV range which translates to a luminosity of (0.8-1.4)$\times 10^{32}$ \ergsec\
(at 3.5 kpc source distance). This fit is displayed in Figure 1b.
The flux and luminosity are (0.8-2.3)$\times 10^{-13}$  \flux\ and (1.1-3.4)$\times 10^{32}$ \ergsec\
in a wider energy band of 0.1-50.0 keV.
These values are similar to the ones derived in Balman (2010)
from the $XMM$-$Newton$  data except for the fact that there is no need for
two different N$_{\rm H}$ to fit the total \cha\ spectrum.
The N$_{\rm H}$ value derived from the fit is consistent with the interstellar values
as in the previous fit. 
The physical interpretation of the two thermal plasma component fit
may be an indication of the multi-temperature nature of the plasma in the boundary layer
or that there are two different contributing X-ray emission regions. This will be
as discussed later.

The combined total spectrum of T Pyx can not be satisfactorily interpreted in the context of 
 a {\it standard} nonmagnetic CV. For
a nonmagnetic CV, the virial
temperature in the inner parts of the accretion disk (kT$_{virial}$=$\mu$m$_p$GM$_{WD}$/3R$_{WD}$)
is kT$_{virial}$=20-29 keV for a 0.7-1.0 M$_{\odot}$ WD (WD mass is 
from Uthas et al. 2010 and Toffelmeier
et al. 2013). The X-ray temperatures on Table 1 show that the flow is already virialized.
A cooling flow-type plasma releases an energy of (5/2)kT$_{max}$
per particle including kinetic and compressional components. The total thermal/kinetic
energy at the inner edge of the disk per particle is (3/2)kT$_{virial}$.
Thus, the plasma maximum temperatures in the cooling flow are limited with
(T$_{max}$/T$_{virial}$$<$3/5) yielding a T$_{max}$ of about 12-18 keV. The X-ray temperature
values (obtained from the total spectrum) of T Pyx are higher than this limit. 
Therefore, the \cha\ observations of T Pyx are not satisfactorily explained
in the context of optically thin boundary layers in quiescent dwarf novae or in the context of 
theoretical expectations from optically thick boundary layers. This will be elaborated
in the Discussion section.

It has been suggested that the source in T Pyx is nonthermal (Scheafer et al. 2013).
I have also fitted a power law model to the total spectrum to confirm such an expectation. 
The spectral parameters of the fit are
an N$_{\rm H}$ fixed at the interstellar value 2.0-3.0$\times 10^{21}$ cm$^{-2}$, a photon index of 
1.6$^{+0.2}_{-0.2}$ and a normalization of
7.6$^{+1.0}_{-1.0}$$\times 10^{-6}$ (\rchisq = 1.71 (dof.17)). The \rchisq\ of the fit is
worse than the fits with the 
thermal plasma models at 98.5 $\%$ confidence level (almost at 3$\sigma$) and can be excluded.
I note that the N$_{\rm H}$ is fixed during the fitting process because it can not be
constrained yielding too low values. This may be a result of the unresolved contribution of line 
emission superimposed over the continuum. In such a case the continuum level is elevated and a simple
continuum model like power-law will still fit the spectrum but at a higher normalization which will
have the effect of lowering the N$_{\rm H}$ parameter in the presence of constant 
interstellar N$_{\rm H}$ absorption. 

The smooth featureless \cha\ spectrum of T Pyx 
hints on plausible non-equilibrium ionization (NEI) plasma effects. A simple model of VNEI in XSPEC
can be used to check this further. 
The model gives acceptable fits with the data only if N$_{\rm H}$ is set free and
yielding an ionization parameter  $\tau$=6.3$^{+3.0}_{-3.0}$$\times 10^{9}$ cm$^{-3}$ s 
($\tau$=nt, n is electron density in this case). Using the normalization of the fit
which is the same as the CEVMKL model, a similar
electron density can be calculated $\sim$ 10$^{13}$ cm$^{-3}$. A double check of these model
parameters yields an ionization timescale of the order of milliseconds and thus a VNEI model
is not physical for this data. 
NEI models like Comptonized plasma models (e.g. CompTT, NthComp in XSPEC) give consistent fits with the 
spectrum of T Pyx only if N$_{\rm H}$ is set free yielding much lower values than the interstellar N$_{\rm H}$.
This problem is removed (fit improved at 3$\sigma$, CompTT is used) if N$_{\rm H}$ is fixed at the 
interstellar value and a second thermal MEKAL model is added to the fit
(a blackbody model is inconsistent) giving a fitted temperature of 0.2-0.55 keV for the MEKAL model
with Comptonized plasma temperatures 3.0-27.0 keV and electron scattering optical depth 2.0-8.0 
(errors are at 90\% confidence level). The temperature of the MEKAL model is very similar to
the first plasma temperature of the double MEKAL model fit in Table 1 (see also Table 2). 
In such case (inclusion of a MEKAL component), a power law model may also yield acceptable fits. 
Since there are no current models
that describe Comptonized plasmas for nonmagnetic CVs (and WDs), no further elaboration will be included, but that
there is consistency with the existing Comptonized plasma emission models.

Note that Selvelli et al. (2008) and Balman (2010) discusses how the blackbody 
model of emission is inconsistent with the data.
A 2$\sigma$ upper limit on the soft X-ray emission from T Pyx calculated from the spectral
fits to the \cha\ data using the
blackbody model is a of kT$_{BB}$$<$25 eV and f$_{BB}$$<$1.5$\times$ 10$^{-12}$ \flux\ in the 0.1-10.0 keV
range. The upper limit on the flux yields a 2$\sigma$ upper limit on the soft X-ray
luminosity of L$_x$ $<$2.0$\times$ 10$^{33}$ \ergsec\ (using 3.5 kpc distance and 0.1-10.0 keV range).

\begin{figure*}
\centerline{
\includegraphics[width=3.0in,height=2.4in,angle=0]{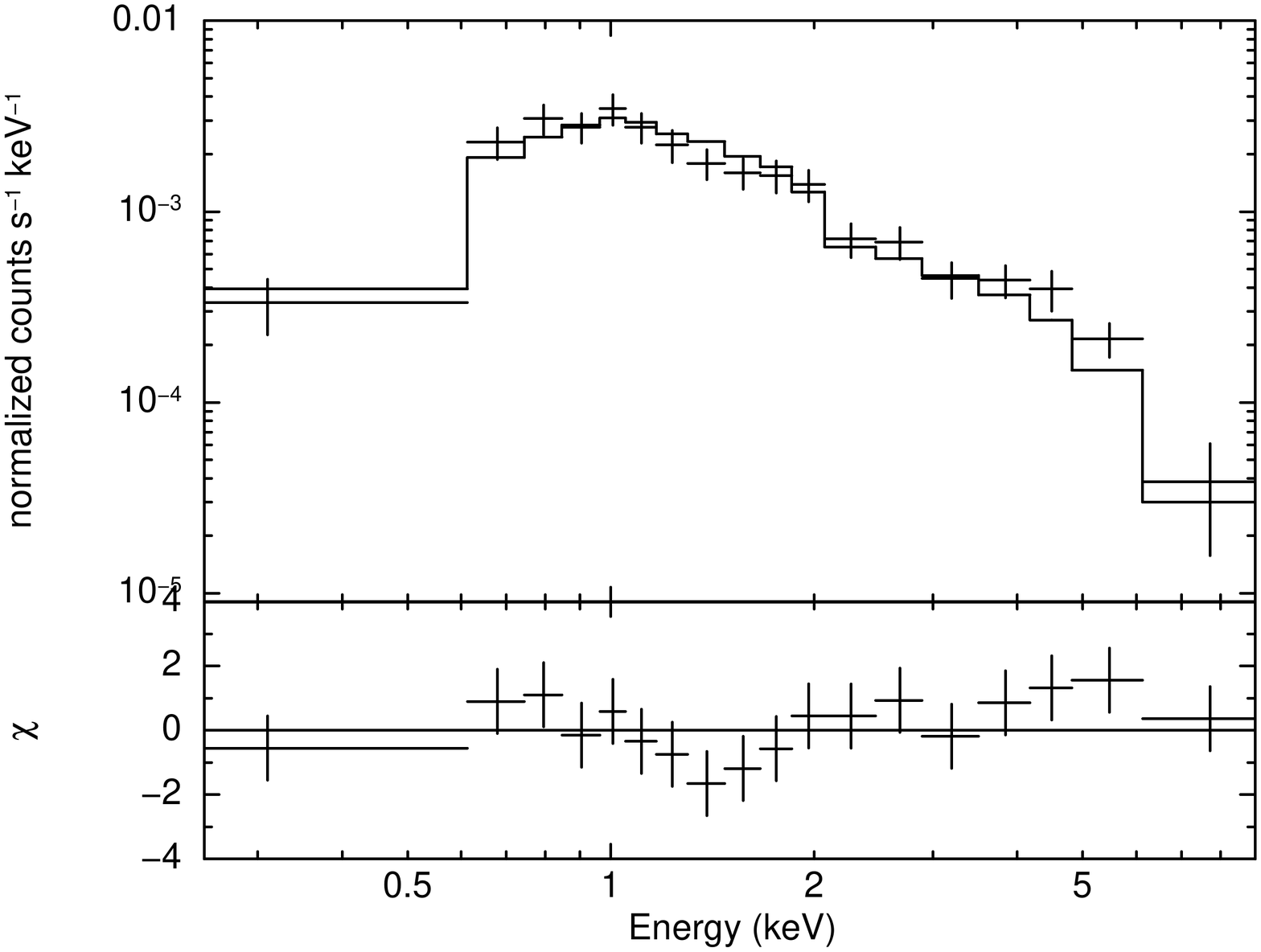}
\includegraphics[width=3.0in,height=2.4in,angle=0]{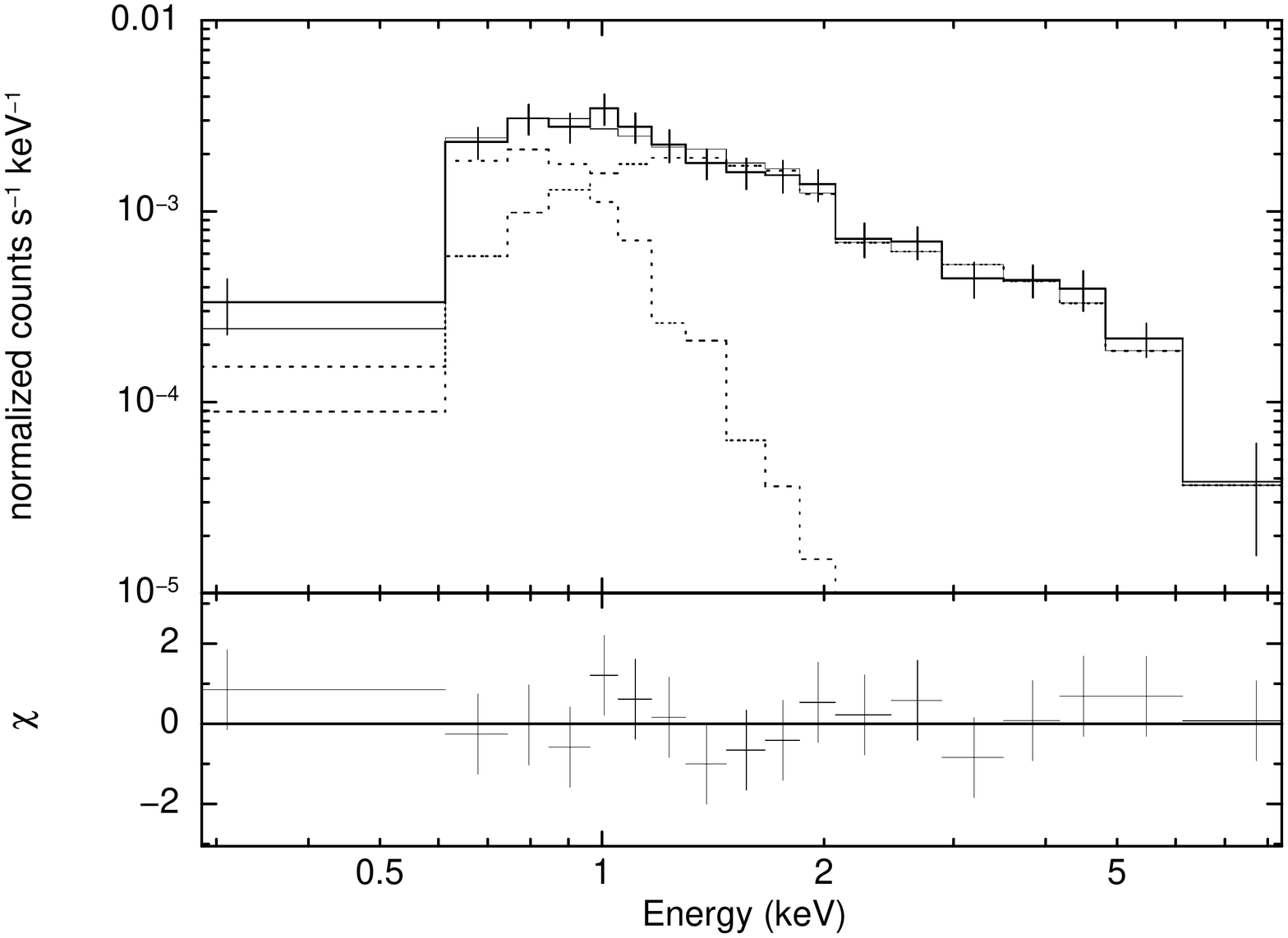}
}
\caption{The figure shows the fitted combined/total \cha\ ACIS-S3 spectrum of T Pyx.
The left hand panel is the fit with the $tbabs$$\times$CEVMKL model and the right hand panel shows
the fitted $tbabs$$\times$(MEKAL+MEKAL) model.
The crosses are the data with errors, solid lines are the fitted model and dotted lines show the contribution of 
the models.
The lower panels show the residuals in standard deviations (in sigma).}
\end{figure*}

\subsection{The Temporal Variations of the Central Source}
\label{subsect:light}

I created background-subtracted light curves with the aid of the CIAO task $dmextract$ to
search for any time variability. The three light curves were,
then, used to generate averaged power spectra (PDS).
I find significant modulation at the binary period of the
system (and its second harmonic) above 99.9$\%$ confidence level
 where 3$\sigma$ power threshold is $>$ 77 taking into account the red noise in the PDS around the
binary period and its harmonic. The binary period is 1.8295(3) hrs; 0.152 mHz (Uthas et al. 2010).
I do not detect any other periodicity.
The PDS with the detected period of 0.155$\pm$0.005 mHz
is displayed in Figure 2, and the folded average light curve is given in the
right hand panel of the same figure.
There is a constant level of emission in the folded
light curve at about $\sim$0.003 c s$^{-1}$.  Superimposed on this constant level, there is
variation between 0.003 and 0.008 c s$^{-1}$ with a mean, $net$, count rate $\sim$0.0025 c s$^{-1}$
yielding a percent modulation in semi-amplitude of about 45\% ([max-min/max+min]$\times$100).

Energy dependence of these orbital modulations are studied using light curves in the 0.2-1.0
and 1.0-2.0 and 2.0-7.0 keV energy bands. Aside from the changing average count rates in the
energy ranges, the modulations and the shape of the average light curve are unaltered. There is
no energy dependence in the orbital modulations in the \cha\ energy band.
T Pyx shows orbital modulations in the X-rays
during the outburst stage with a slightly different shape of the (folded) mean light curve 
(Tofflemire et al. 2013).

\begin{figure*}
\centerline{
\includegraphics[width=4.2in,height=2.3in,angle=0]{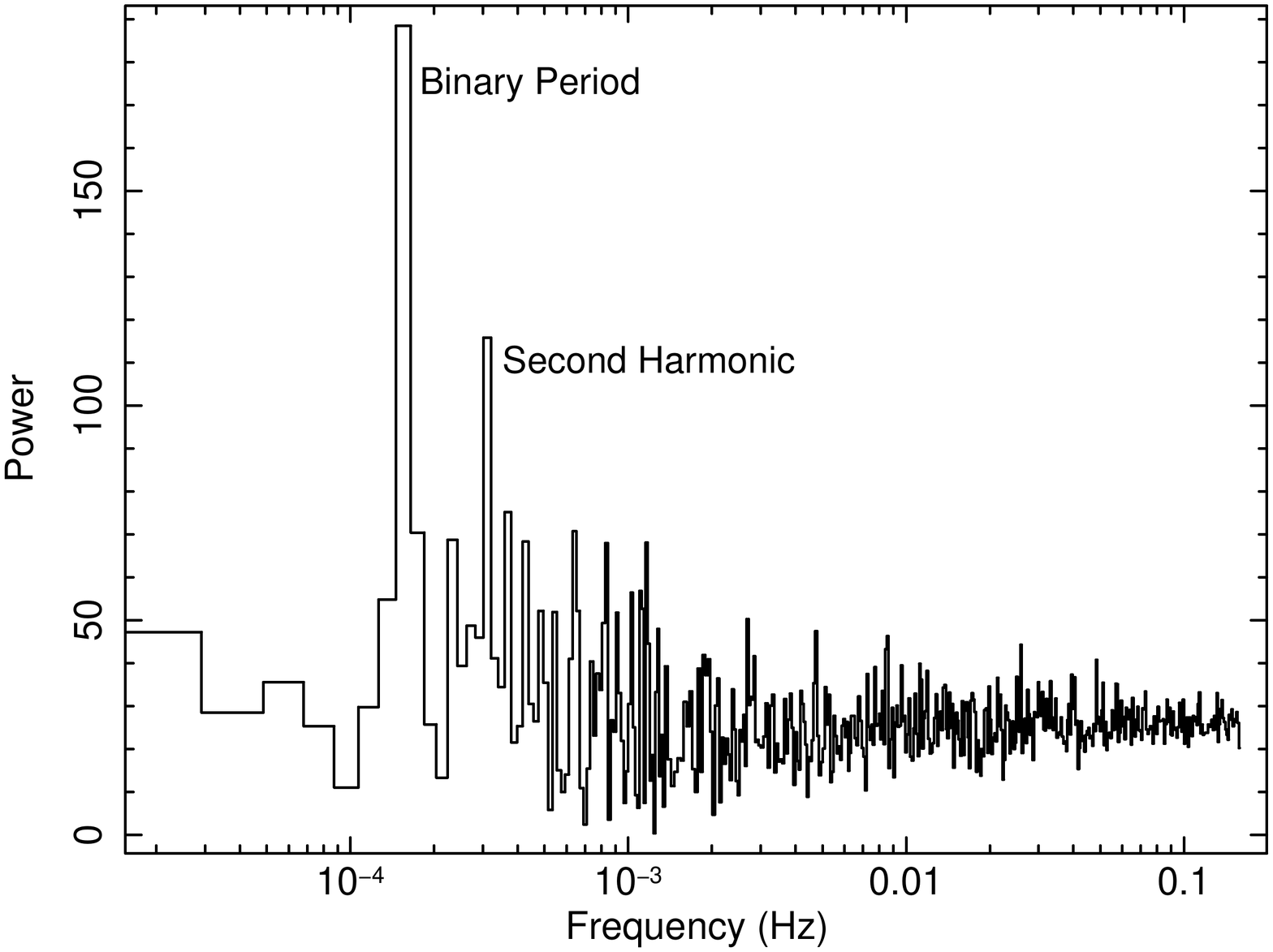}
\includegraphics[width=2.3in,height=2.7in,angle=90]{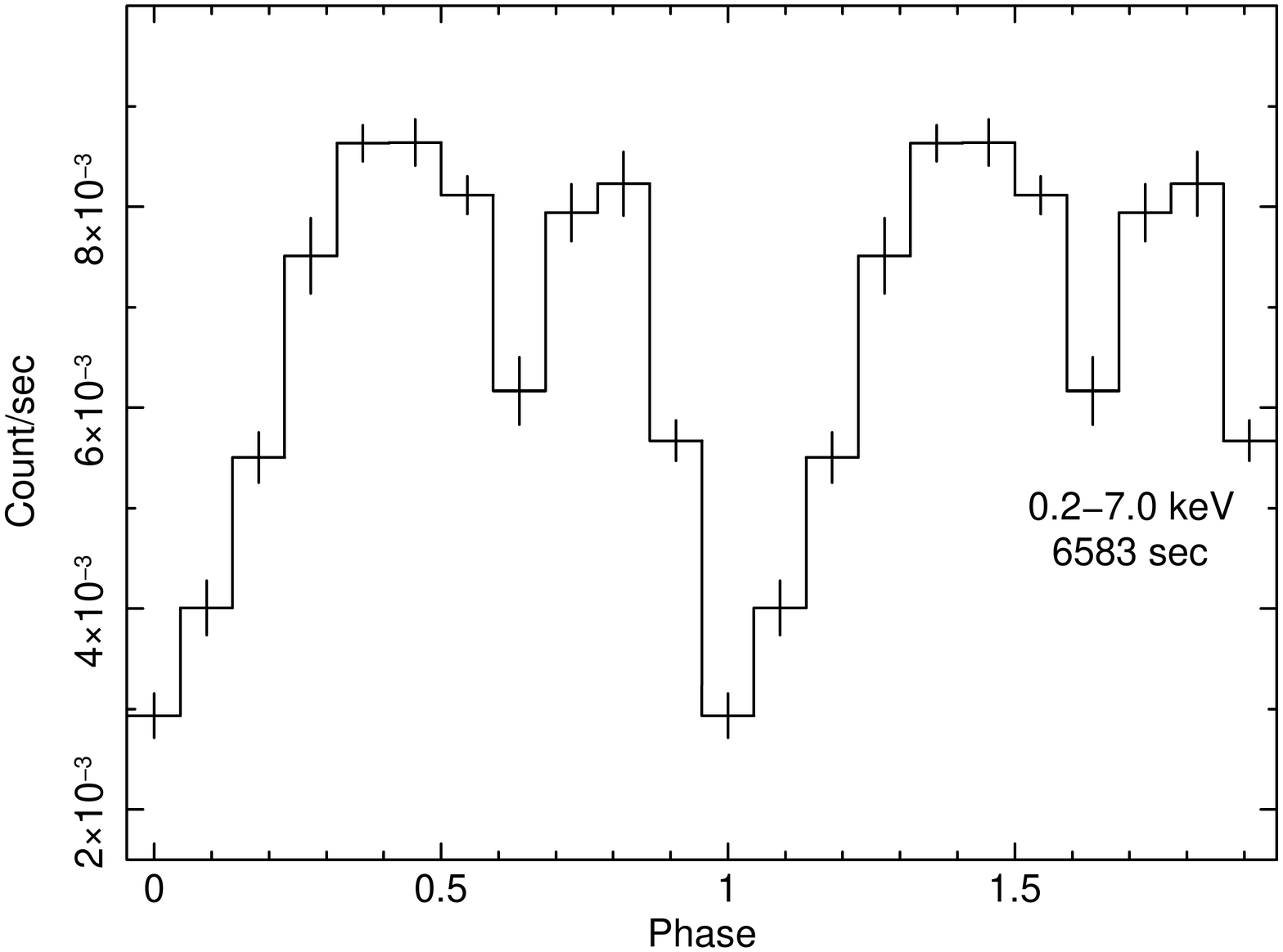}}
\caption{The left hand panel shows the PSD obtained from averaged power spectra of the three observations.
The right hand panel is the
folded average X-ray light curve of T Pyx using the detected binary period.}
\label{fig:light}
\end{figure*}

\subsection{On the Extended Emission around T Pyx}
\label{subsect:ima}

A low S/N ($\sim$ 4-5$\sigma$) extended excess emission was recovered in the 
$XMM$-$Newton$ data of T Pyx (Balman 2010).
In order to assess this better a long observation of T Pyx was obtained in quiescence
with a much better spatial resolution capability.
To improve the spectral statistical quality and effectively recover
the extended emission using deeper imaging,
the three \cha\ observations of T Pyx (see section 2)
were merged with the standard procedures (see http://cxc.harvard.edu/ciao/threads/combine/).
A specific new tangent point (RA-DEC: 09 04 41.49, -32 22 47.65) is used to
match the three OBSIDs which projects the events to the nominal point of ACIS-S (x=4096.5, y=4096.5 in
WCS--World Coordinate System) for all three observations in the process of the analysis. Thus,
the events in the three observations have been reprojected to this new tangent point and the
analysis is performed at/around the nominal point of ACIS-S.
During any of the re-projection processes in my analysis
pixel randomization has been turned off. In addition, all events files have been re-processed
(using {\it acis-process-events}) without
randomization of any kind (pix-adj=NONE, yielding original position of detected photons).
In general, I have analyzed the T Pyx data originally generated in 2011 (processed with pixel 
adjustment set to
randomize using 0.5 pix) and archived in 2012 (processed with pixel adjustment set to EDSER,
a new event repositioning algorithm to improve pixel resolution) and found similar results 
(using pix-adj=NONE).
In merging events, there are two important factors, one is the fine astrometric shifts that are
necessary to match sources, the second is the choice of a common tangent plane as used in this
analysis with the assumed new RA-DEC position to project all the events files.
To improve on any systematic shift in the coordinate systems between observations
and enable usage for  high-resolution imaging at the sub-pixel level of the combined data sets,
the aspect solutions of the observations have to be corrected to make it consistent with
each other. Certain astrometric shifts were determined for the correction of the aspect solutions and
applied as described in the analysis threads, updating WCS and
re-projecting the events matching them to the second observation of T Pyx.
In this analysis, shifts were calculated using the projected events files
to the nominal point of ACIS-S and also checked with the shifts obtained from the
original individul data sets. The results show that the shifts in the x and y axis
for the first and the third observations are in a range $\le$\ $\pm$0.5 pixels including errors.
The range is a result of different radii chosen
for {\it dmstat} to calculate centroids for the source yielding
slightly different shifts for different radii from a given observation.
The shifts used in this study are x=0.5 pix,y=-0.07 pix for the first observation
(ID 12399) and x=0.04 pix,y=0.03 pix for the third observation (ID 13224).
Note that the shift variations maybe because the source profile is
imperfect due to the superimposed extended emission with a different symmetry
(see discussions in the following paragraphs).
Using these astrometric shifts obtained from a given source radii (5-7 pix) yields similar
structure at around the nominal point, however slightly defocused (slightly unmatched). 
I have tested the data for several
sets of shifts and using  small extra systematic shifts in the given range of shifts in pixels
keeping track of bright parts and the elongation towards south
in the individual images to match them in the final merged image. The shifts given above
for this study are final shifts used in the analysis.
Mainly, one needs to do only one re-projection of the individual events files to the
nominal point using a new tangent point (RA-DEC) as one starts the analysis with and
next apply astrometric shifts to match these events files. In general, all through out the
analysis the location of the nominal point and the chosen RA-DEC position should remain intact.
If any slight miss-match occur, in this case,
it is suggested to do a second re-projection of shifted events using the event file
of the observation chosen as reference for the other two observations (second observation
in my case). This process corrects the match between sky coordinates and  WCS and thus,
the location of the nominal point (and put the data on common tangent planes).
However, it leaves the profile of the source
unaltered from the profile with the inadequate shifts and the incorrect aspect solution
calculated with these inadequate shifts. Therefore, care must be given to check allignment
at any step of the analysis.

To create the PSF, the CHART ray tracer was utilized using the necessary off axis angles 
(taken from the merged
event file), the combined source spectrum
(see section 3.2) and the entire exposure time (98.8 ksec). Next, MARX version 4.5.0 was
used to project the PSF rays onto the detector plane and finally an events file
was generated for the source PSF. 
A single PSF was generated for the analysis
to ensure no centroiding/allignment problems for the PSF.
The three observations were obtained within one week time and a single PSF, thus is suitable.
In addition, all three observations are pointed close-by and
have no off-axis angles larger than 20 arc sec where the standard PSF is mostly unaltered.
Since the combined source spectrum, thus source flux is directly used
with the exposure time, the created PSF is expected to be very similar to the combined/merged
\cha\ observation of the source.
For the analysis several PSFs were generated and inspected.
It has been calculated and checked that in the same size region at round the PSF kernel size 
both the simulated PSF data and the source data images yield very similar photon counts
around 510$\pm$15. This means that no further secondary normalization can be made
on the PSF without changing the constant source flux that is in the data and the simulated data (PSF).
If there is a necessity, a new PSF should be recalculated instead of a secondary normalization
for simulated PSF and source data correspondance.
Moreover, images derived from these two event files (source data and simulated PSF data)
have to be treated exactly the same way in any analysis so that this correspondance of flux/counts
remain the same.
In addition, I stress that PSF events files have no aspect solutions that can be properly used for
astrometric corrections to match and use particularly with high-resolution imaging analysis at the 
sub-pixel level, 
as in the standard events files.
Therefore, merging PSFs needs to be cautioned for sub-pixel imaging analysis. Also, note that
in this analysis the individual PSF events files have to be projected to the nominal point
for any subtraction process where aspect solution files will be necessary.

Since the standard \cha\ pixel resolution did not properly resolve the structure of the
extended emission (see the discussion of the 1-D radial profiles later in this section),
a 0.1-pixel scale (corresponding to a  0.\arcsec05 angular size) 
was used to push  the spatial resolution of the data close to the instrumental limit for a
detailed imaging analysis.
I note here that the photon positioning capability of \cha\ is good,
but the caveats due to the detector, HRMA and aspect errors 
results in a sub-structural resolution of \cha\ ACIS-S no better than 0.\arcsec2-0.\arcsec3 .
Therefore, structure sizes smaller than this can not be resolved.
Moreover, I stress that there is no study or known calibration work done to test the sub-pixel
response of the PSF at off-axis angles (private communication with the Chandra Calibration team through 
Help Desk).
Therefore, it is most suitable to do the analysis at around the
nominal point of ACIS-S where resolution is best, well known and well calibrated.
Then, the choice of merging the three events files of T Pyx via a new tangent point RA-DEC that
projects the events to the nominal point is the correct choice at hand.

During the high-resolution imaging analysis of T Pyx at the sub-pixel level,
a slow PSF subtraction methodology is used. The data image and the PSF image
created with the same 0.1 ACIS pixel scale in the 0.2-7.0 keV band with matching coordinates were subtracted 
in iterative steps. First, with no smoothing applied
before subtraction, then
with  1$\times$1 pixel smoothing applied before subtraction,
finally  2$\times$2 pixel smoothing applied before subtraction while a persistent structure
remaining in the PSF-subtracted data images is explored. A Gaussian smoothing is used
with the task $csmooth$. Note that the task $csmooth$ (an adaptive smoothing technique) 
is run with constant/same minimum and maximum smoothing pixels. Also, non-integer values of
pixel smoothing has been used to check how well the iterative process works. 

The resulting image after PSF-subtraction (using the 1$\times$1 pixel smoothing) was further
smoothed by 2$\times$2 pixels yielding a spatial resolution of 0.\arcsec25 and
displayed in Figure 3 (right hand panel). 
The left hand panel of Figure 3 shows the 15 arscec vicinity of T Pyx, and the
the middle panel shows the unsubtracted sub-image derived from the merged events file. 
Note here that, smoothing decreases
spatial resolution in the expense of signal-to-noise 
since it re-evaluates the data values in the images at given sub-pixel 
positions and increasing smoothing
makes the resolution closer to the standard pixel images of \cha\ ACIS-S. 

The PSF-subtracted final image indicates some extended emission with an elliptical
shape.  The  outer semi-major axis is $\sim$
1.\arcsec0  and the outer semi-minor axis 
is $\sim$ 0.\arcsec5\ . I underline that the subtraction process
also yields a depression of counts at the center, which is expected if the
PSF and the data image is correctly normalized, size of this depression in counts is about
0\arcsec.2-0\arcsec.4 at the sub-structural resolution limit.
This possibly indicates a torus-like or ring-like  structure around the nova.
In addition, there seems to be an elongation towards the south to about 1\arcsec.85 from the point source. 
The southern region of extension
comprises about 13\% of the extracted X-ray nebulosity.
I checked the consistency of the X-ray nebula by analyzing the
three datasets of T Pyx and subtracting an appropriate PSF from each individual image. 
Though the S/N is low,
structures resembling  Figure 3 were seen. 
In addition, I checked the given PSF subtraction procedure
with some weak sources that show no pile-up and found no  excess structure as in T Pyx.
My experience in these trials has been that the PSF image clears the source data image more or less using 
1$\times$1 pixel smoothing before subtraction (at 0.1 ACIS pixel scale) leaving 
behind no definite structure but few scattered spots.

I determined an approximate nebular and point source count rate by using the PSF-subtracted 
total image (merged) of the nebula within about a radius of 5\arcsec 
 (the representative extraction area shown in Figure 3a--large circle), since
it is about the size of the
optical remnant (see Schaefer et al. 2010) and the \cha\ nebulosity is within 2\arcsec\ of the point source.
As in the imaging analysis  discussed/outlined above, 
I used either no smoothing or 1$\times$1 pixel and 2$\times$2 pixel
smoothing for both the PSF and the merged source image
with the 0.1 ACIS pixel scale to obtain an approximate nebular count rate. 
I used a similar size photon extraction region obtained elsewhere in the image
to account for the number of background photons 
(the representative extraction area shown in Figure 3a--large circle).
This analysis yielded an approximate count rate of (0.0023$\pm$0.0010) c s$^{-1}$ (about 40\% error) for the
extended emission and
$\sim$ 0.003 c s$^{-1}$ for the central source over the 0.2-7.0 keV energy range
(the rates indicate no concerns for pile-up).
Because of the caveats of the high-resolution imaging  
analysis at the sub-pixel level, it is most likely that the lower
half of the count rate range, thus $\sim$ (0.0013-0.0025) c s$^{-1}$ is plausibly more representative of the  
X-ray nebular emission.
The signal-to-noise ratio of the extracted
nebulosity is between 6-15
using the entire range of the X-ray nebular count rate ($\sim$ 6-10 for the lower half of the count rate
that is a better estimate of the S/N)
( S/N=E/$\sqrt{(B+P)}$; E is the net nebular counts, B is the background counts
and P is the counts in the PSF).

In order to inspect the morphology of this nebulosity
radial profiles of T Pyx are calculated in different directions and displayed
together  with the radial profile of the simulated PSF used for the analysis in Figure 4. 
For the analysis, 
the task  $dmextract$  is used along with the circular annuli within a region of 
about 3\arcsec\ radius to
indicate the background (Gaussian errors are assumed in the calculations).
The total number of photons are about the same in the source and PSF profiles.
The T Pyx radial profiles are created from a sector with a narrow opening angle of
30 degrees centered on the east, west,
south and north.
A radial profile of the ACIS-S3 PSF with about the same number of photons as in the source radial profile 
calculated from the same sector with the given directions is included in
all the figures. Gaussian errors are assumed. 
The extended nebulosity in the western part is evident. 
Moreover, the northern profile is narrower than the PSF indicating that the underlying central point
source has less emission and thus consistent with a smaller PSF.
The southern profile is larger in extend than the
PSF indicating the existence of the southern elongation as shown in Figure 3.
The excess in the central part of the southern radial profile is consistent with the excess
in the northern sector and is originating from the superimposed elliptical nebulosity
as projected mainly on the minor axis.
Therefore, one needs to be very careful in
interpreting circularly
symmetric radial profiles where superimposed non-circular and/or local geometries may not
show explicitly in a total 1-D circular radial profile.
In order to show the compliance of the T Pyx radial profiles with point source PSF
with less counts than the total counts  detected from T Pyx, I prepared
similar figures using the same source radial profiles from a sector of 30 degrees,
but this time overlayed
point source PSF with counts in the range of 250-300.
Figure 5 shows the correspondance with total counts from a smaller point source.
Note that particularly the eastern and northern radial profiles yield consistent results with
a smaller point source size.
There is an excess superimposed on the central regions
in all the figures in Figure 5 (and Figure 4). I suggest that this may result from
an elliptical nebulosity
superimposed on the semi-minor axis ($\sim$ 0.\arcsec5).

An Astronomical Telegram on this \cha\ data claims that the extended emission is due to an
HRMA artifact (Montez et al. 2012). The HRMA artifacts
for weak sources like T Pyx are less than 5$\%$ of the total emission (private communication
with the CXC calibration team through the Helpdesk) and are from a very localized
region of  0\arcsec.6 by 0\arcsec.4 (see  http://cxc.harvard.edu/ciao/caveats/psf-artifact.html).
The entire size of the X-ray nebula, as I have derived, is much larger than an artifact zone
and the nebular emission is about or more than 30$\%$ of the total emission from the source
showing that it is not an HRMA artifact. Moreover, note that the extension towards the south
is also about 13$\%$ of the total source emission.  The artifacts should be looked for in  the original data and images
and not in the images from reprojected and merged events files as in
Montez et al. (2012) since WCS is also updated during the projection process.
Reprojection and merging of events do not create HRMA artifacts. Analyses of the original data sets
(using CIAO task $make$-$psf$-$asymmetry$-$region$) yield 
zero photons in the first data set, six photons in the second dataset and seven photons
in the third data set of T Pyx, that operlap with the HRMA artifact regions which totals to 16 photons
that may be associated with an artifact.
Thus, there are no concerns from significant contribution of HRMA artifacts into the nebulosity. 
Moreover, the radial profile presented in this ATEL is in "counts" and shows definite misallignment.
The unit "counts" is not surface brightness in  "counts/pixel$^2$" and 
the extraction area increases as you go from the center of the PSF/source to the wings of the profiles. 
Therefore, there is a normalization problem and a definite 0.3-0.4 arc sec positional centroid mismatch between 
the merged PSF and the merged T Pyx profile in Montez et al. (2012) which does not exist in Figures 4 and 5
in this paper. 

\begin{figure*}
\centerline{
\includegraphics[width=2.5in,height=2.6in]{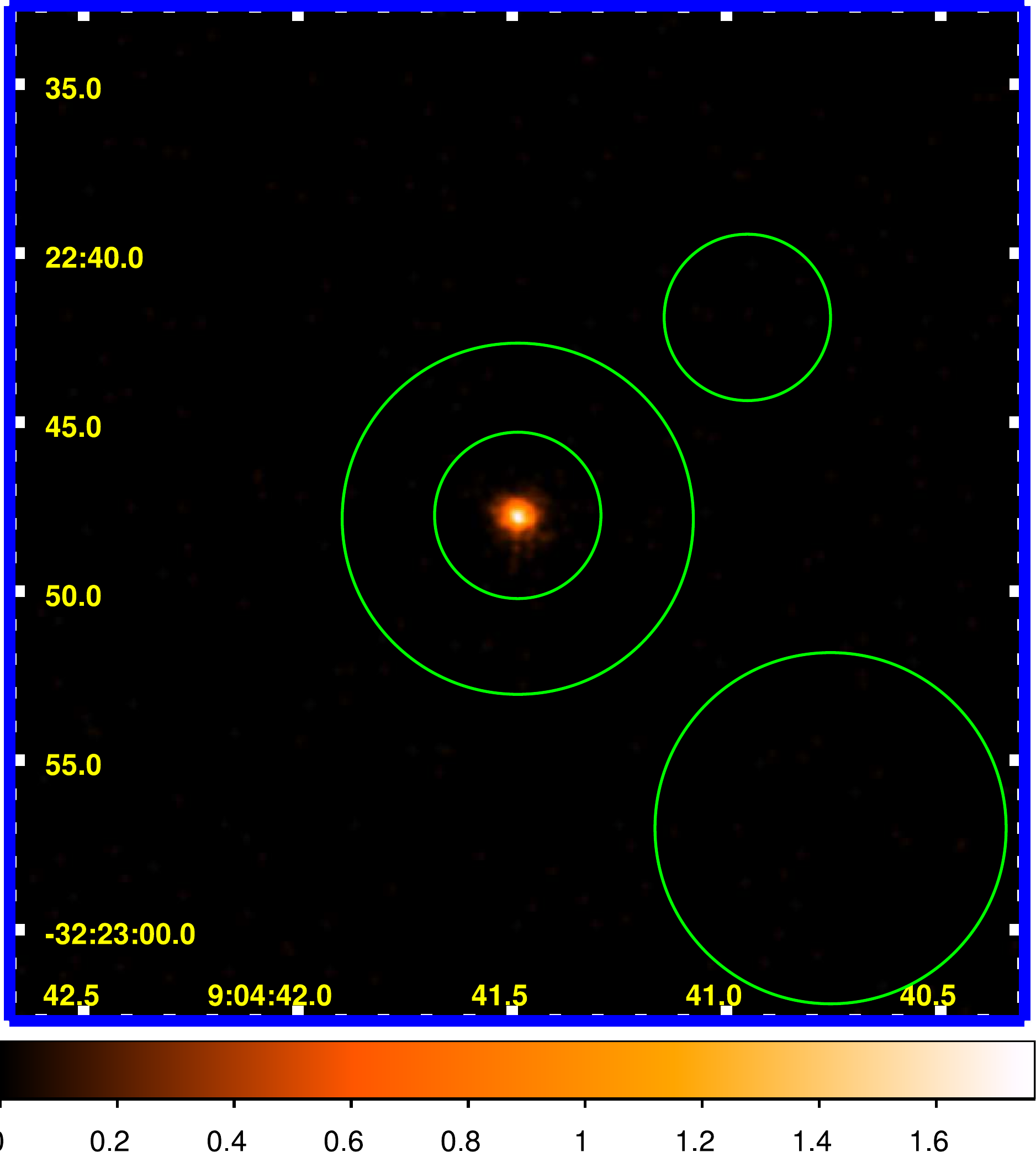}
\includegraphics[width=2.5in,height=2.6in]{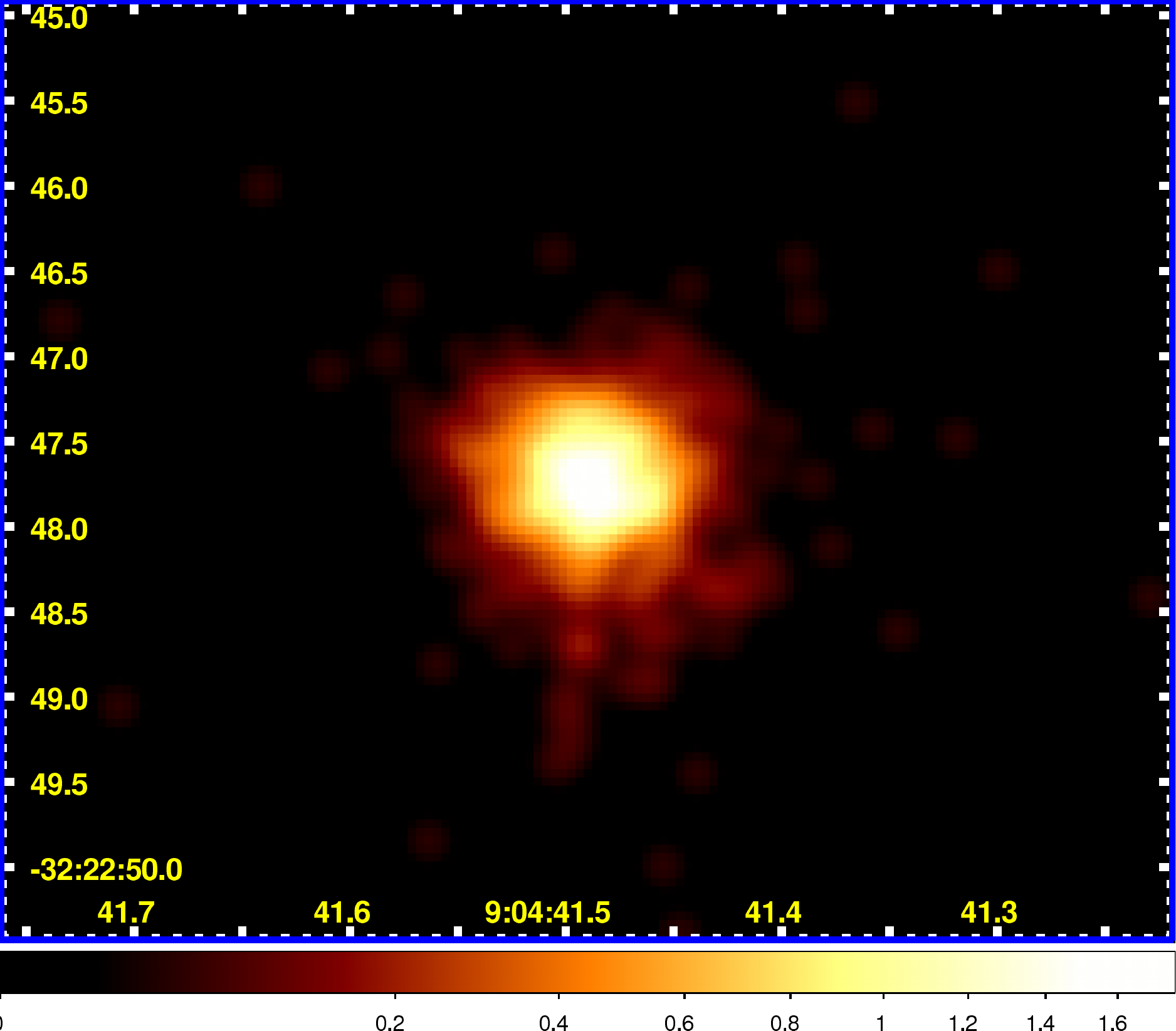}
\includegraphics[width=2.5in,height=2.6in]{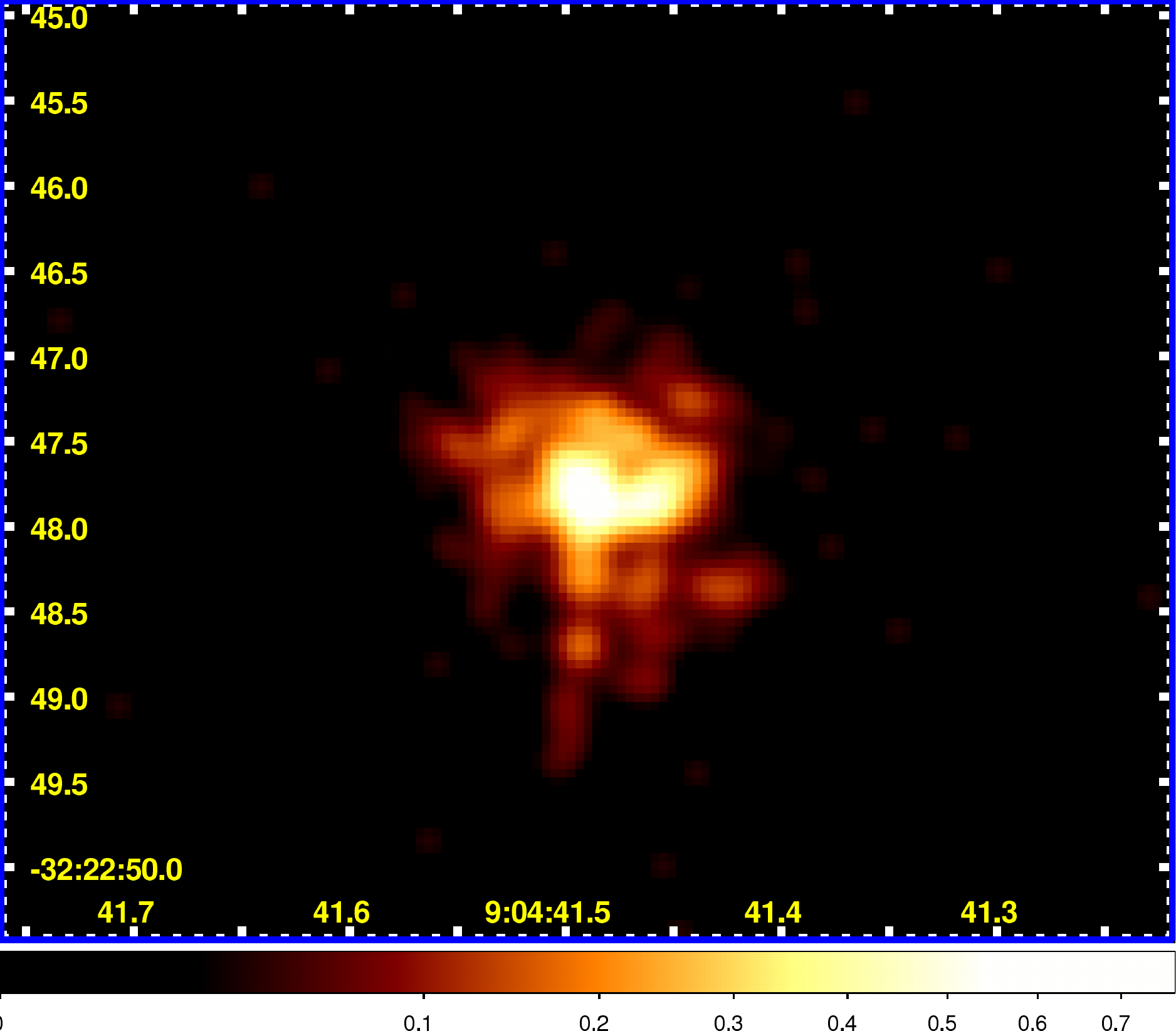}}
\caption{X-ray images of T Pyx  Nebula in the
0.2 to 7.0 keV range.  North is up and West is to the right.
The left hand panel shows the vicinity of the source within 15 arcsec radius and the photon extraction radii 
for the source and background are overlaid for 5 and 2.5 arcsecs. The
middle panel is the image without the subtraction of the central source PSF. The
right hand panel is the PSF-subtracted image. 
The resolution is 0.$^{\prime\prime}$25 per pixel in the right hand panel.
The axes on the figures show RA (x-axis) and
DEC (y-axis). The images utilize different brightness levels using the
squate root scaling. }
\label{fig:ima}
\end{figure*}

\begin{figure*}[ht]
\centerline{
\includegraphics[width=1.95in,height=1.85in,angle=90]{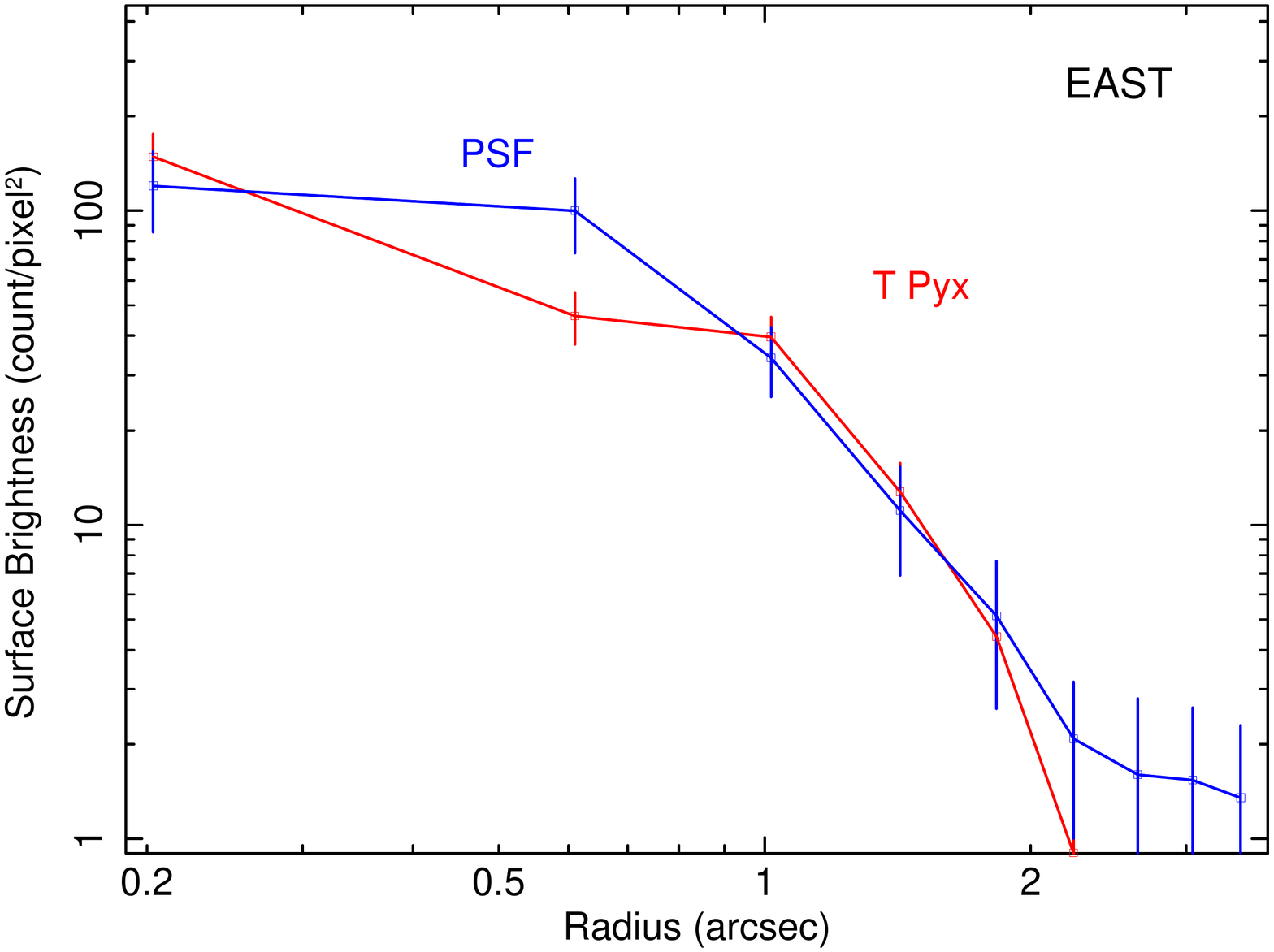}
\includegraphics[width=1.95in,height=1.85in,angle=90]{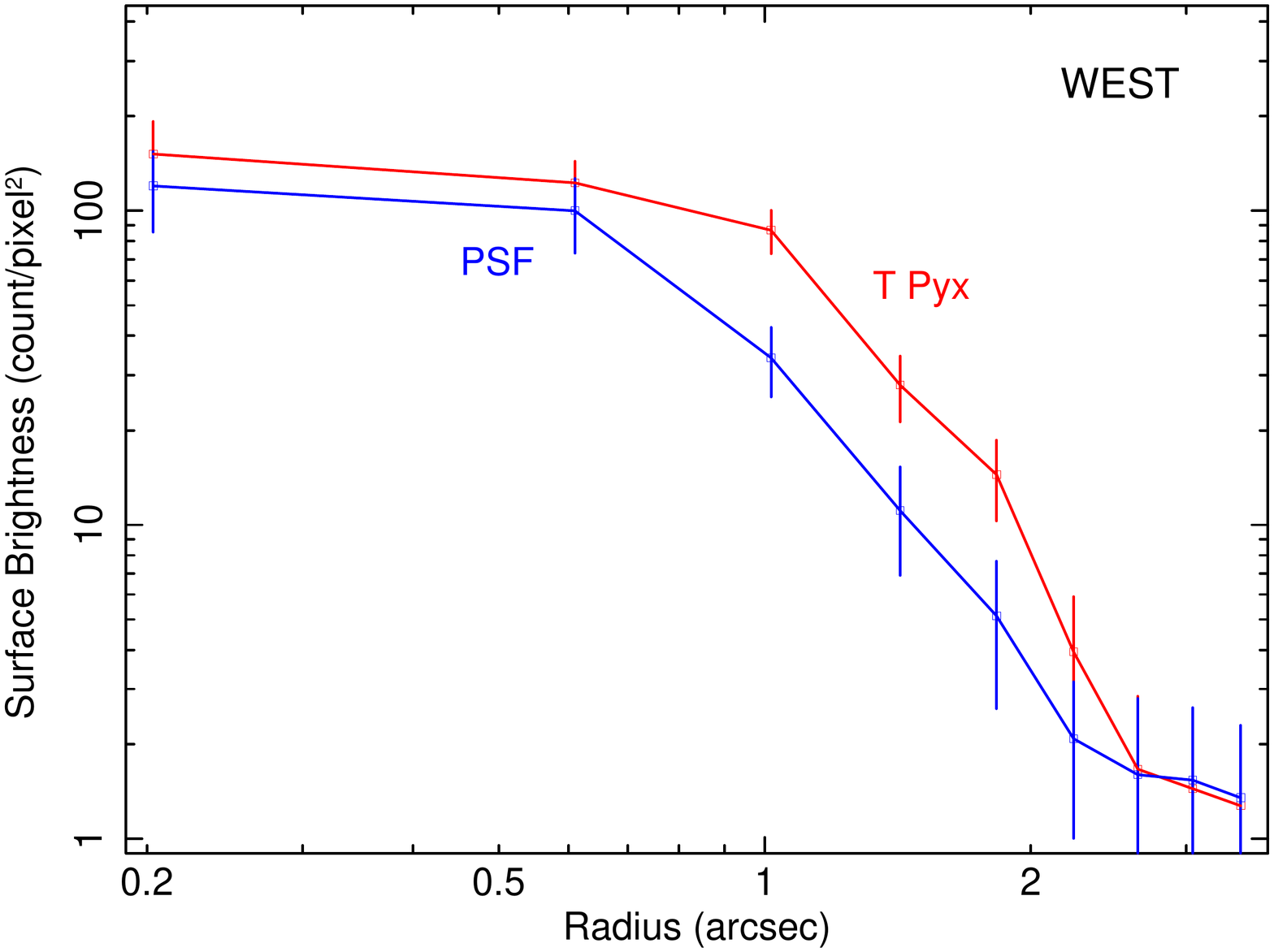}
\includegraphics[width=1.95in,height=1.85in,angle=90]{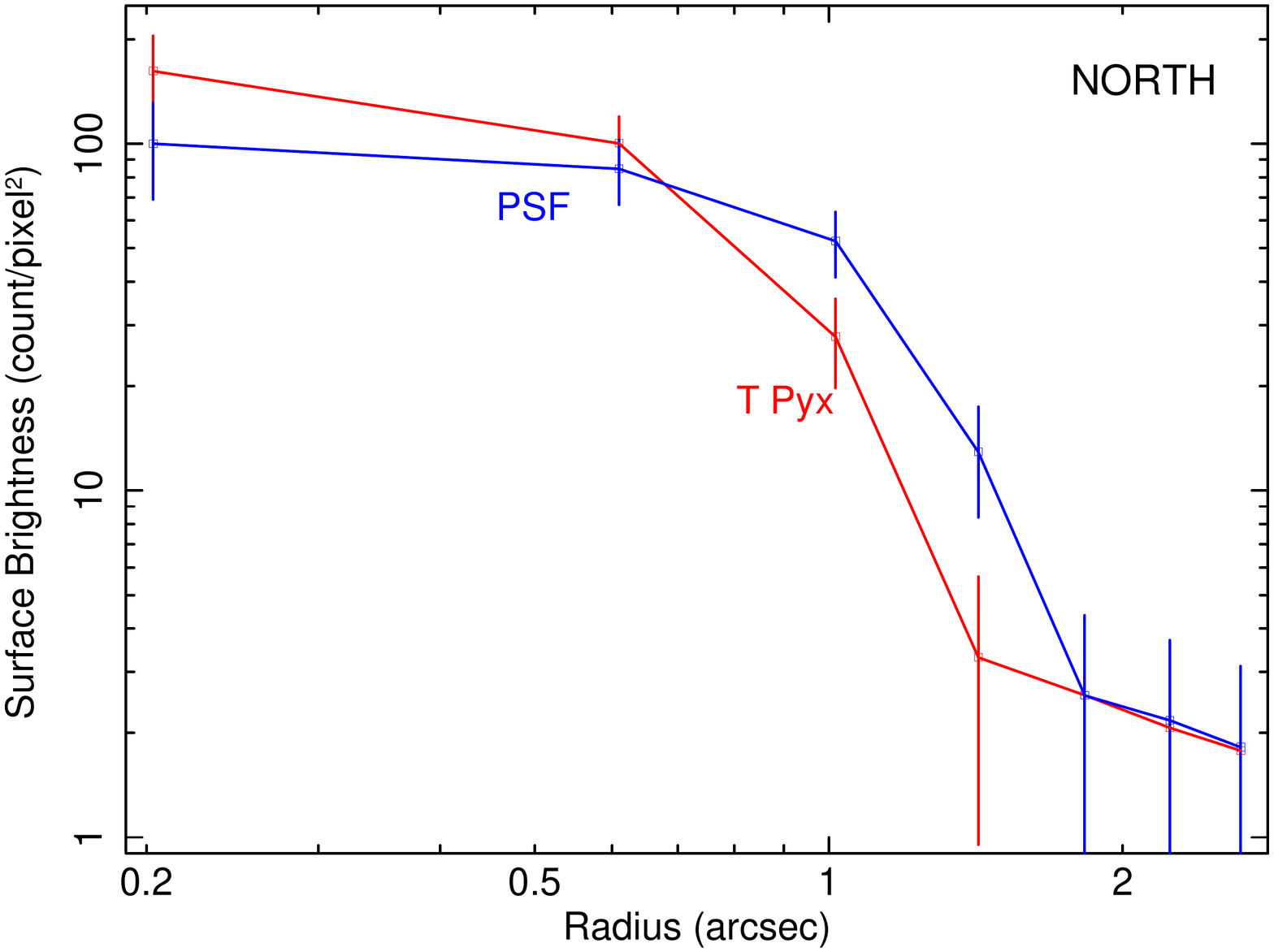}
\includegraphics[width=1.95in,height=1.85in,angle=90]{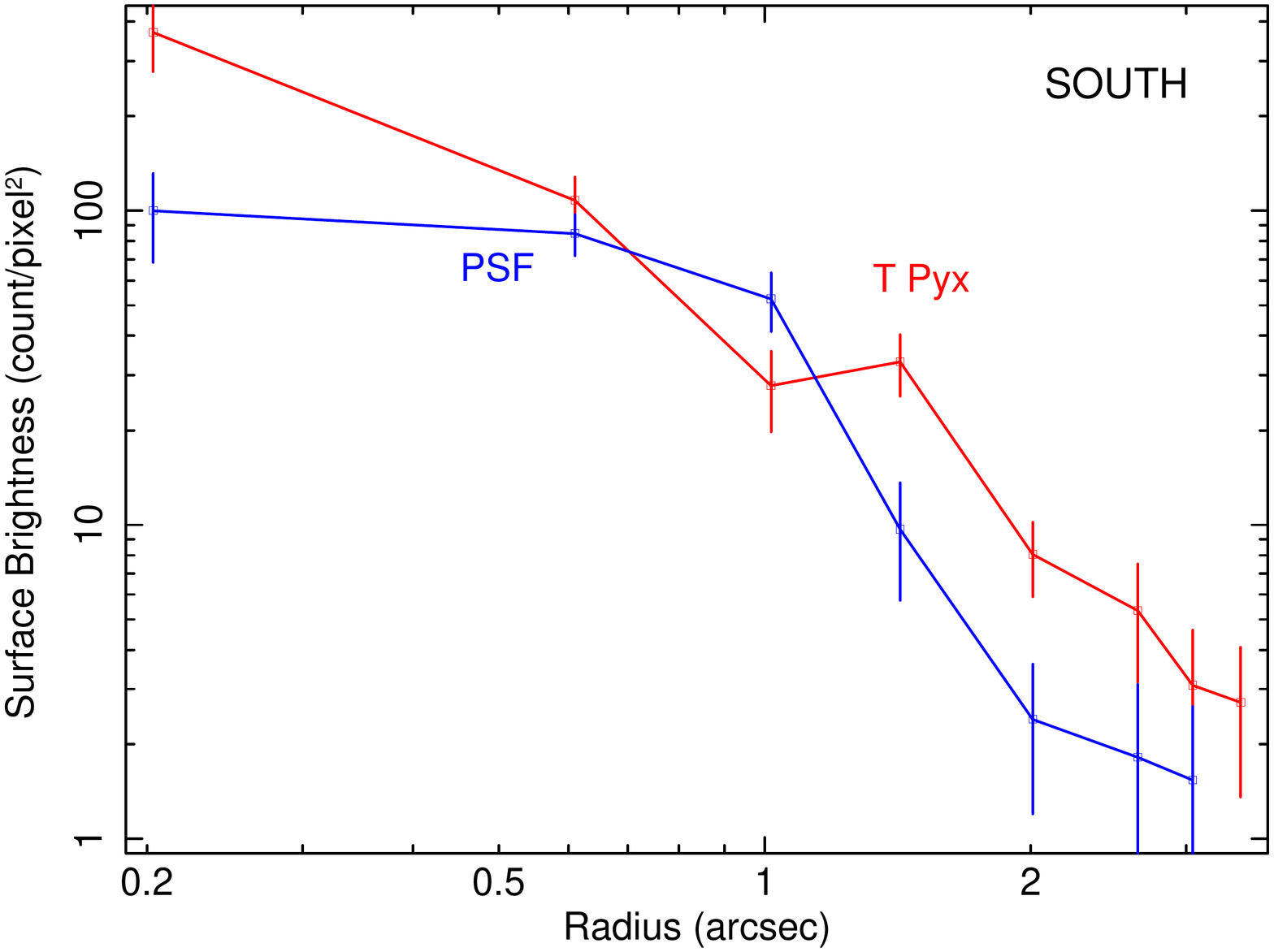}}
\caption{
The radial surface brightness profiles of the source T Pyx
extracted from a sector-area of opening angle 30 degrees.
From the left hand to the right hand panel the profiles are in 
the eastern, western, northern, and southern
directions, respectively. 
In all panels, an equivalent \cha\ ACIS-S3
PSF radial profile of about 500 photons calculated from the same sector is overplotted.
Gaussian errors are assumed.}
\end{figure*}

\subsection{A Method for Deconvolution of the \cha\ Spectrum of the X-ray Nebula and the Central Source}

The excess nebulosity around T Pyx and the inadequate results from the analysis
of the total/combined spectrum possibly support the existence of two different origins of
emission within the
total spectrum; one being the CV which is the point source and the other the X-ray nebula.
The nebular emission is less than 2.\arcsec5 around the point source (about 1\arcsec\ radius)
and it is weak compared to the point source. As a result of difficulty of extracting
photons from separate regions to collect nebular photons, and source photons, I used the
technique described below to deconvolve the two spectra:
I have assumed 16 equal energy channels between 1-1024 channels of the \cha\ ACIS-S
(e.g., 1-63,64-127,128-191\dots\ 512-575). 
For each of the given 16 energy channel ranges, sub-images were created using the merged events file
and the PSF events file as used in the imaging analysis of the previous section.
The PSF and the source images created at the given energy channel ranges
were smoothed by 2$\times$2, or 3$\times$3 pixels where each pixel was at 0.1 ACIS pixel scale
(using a Gaussian kernel).
Next, PSF images were subtracted from the source images using  the same ACIS pixel scale, smoothing, and matching
coordinates.
Resulting PSF-subtracted images were thresholded to have zero
counts for the negative pixel values. Using these subtracted images, the number of counts in each channel
range (representing the excess nebular counts) were
calculated utilizing the same photon extraction radius
as the combined source spectrum (either 5\arcsec radius with the 2$\times$2 pixel smoothed images
or 2.\arcsec5 radius with 3$\times$3 pixel smoothed images have been used). The 5\arcsec and the 2.\arcsec5 photon
extraction radii for the source and background is displayed in Figure3a.\
The nebular spectrum was created by replacing the counts in the re-binned combined spectrum
(re-binned to have 16 channels) created using 5\arcsec\ or 2.\arcsec5 photon
extraction radius. The same
background spectrum is used for the nebular spectrum as derived for the combined total spectrum
assuming the corresponding photon extraction radius. Figure 6, left hand panel, shows the nebular spectrum
with the 2$\times$2 pixel smoothed image subtraction in comparison with the total combined spectrum. The
middle panel shows both of the nebular spectrum that uses the 2$\times$2 pixel smoothed image subtraction (in black)
and 3$\times$3 pixel smoothed image subtraction (in red).

The deconvolution method used here has some caveats. The first one being the low counts
in the selected energy channels which may result in appreciable mismatch between PSF and source photon
positions within the PSF kernel. The choice of background and extraction radius 
may also yield slight variations of the calculated nebular spectrum.
As described above, to mitigate this, 2$\times$2 pixel (0.\arcsec25 resolution), or 3$\times$3 
pixel (0.\arcsec35 resolution--larger than half the standard \cha\ pixel) 
Gaussian smoothing has been used for the imaging analysis. It is evident that similar
spectra is derived, but, I underline that
as the counts get lower in the harder energy channel ranges,
the image subtraction is relatively less reliable. On the other hand,
I stress that completely smoothed out images has been subtracted
in all the energy channel ranges.

\begin{figure*}[ht]
\centerline{
\includegraphics[width=1.85in,height=1.95in,angle=0]{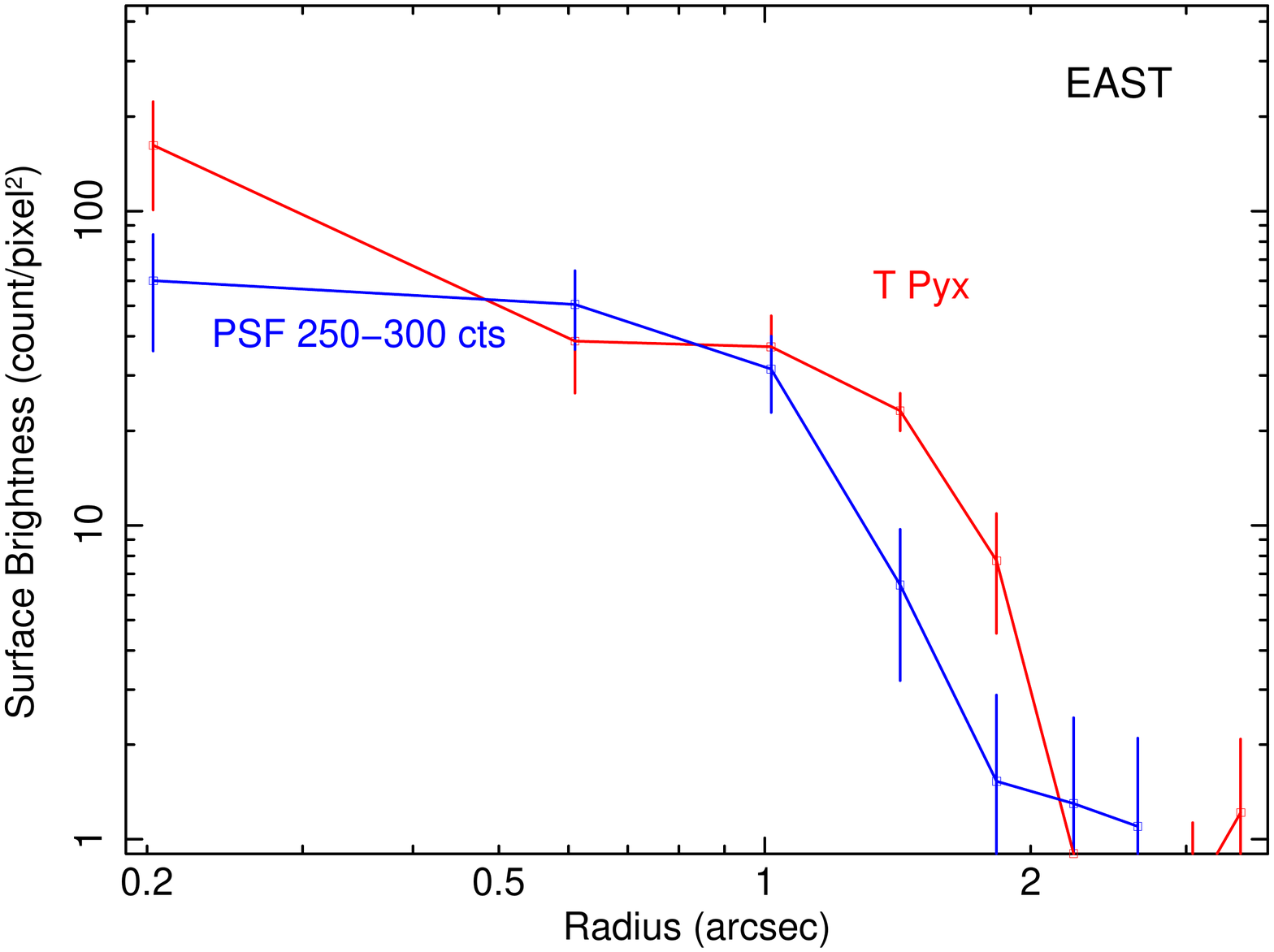}
\includegraphics[width=1.85in,height=1.95in,angle=0]{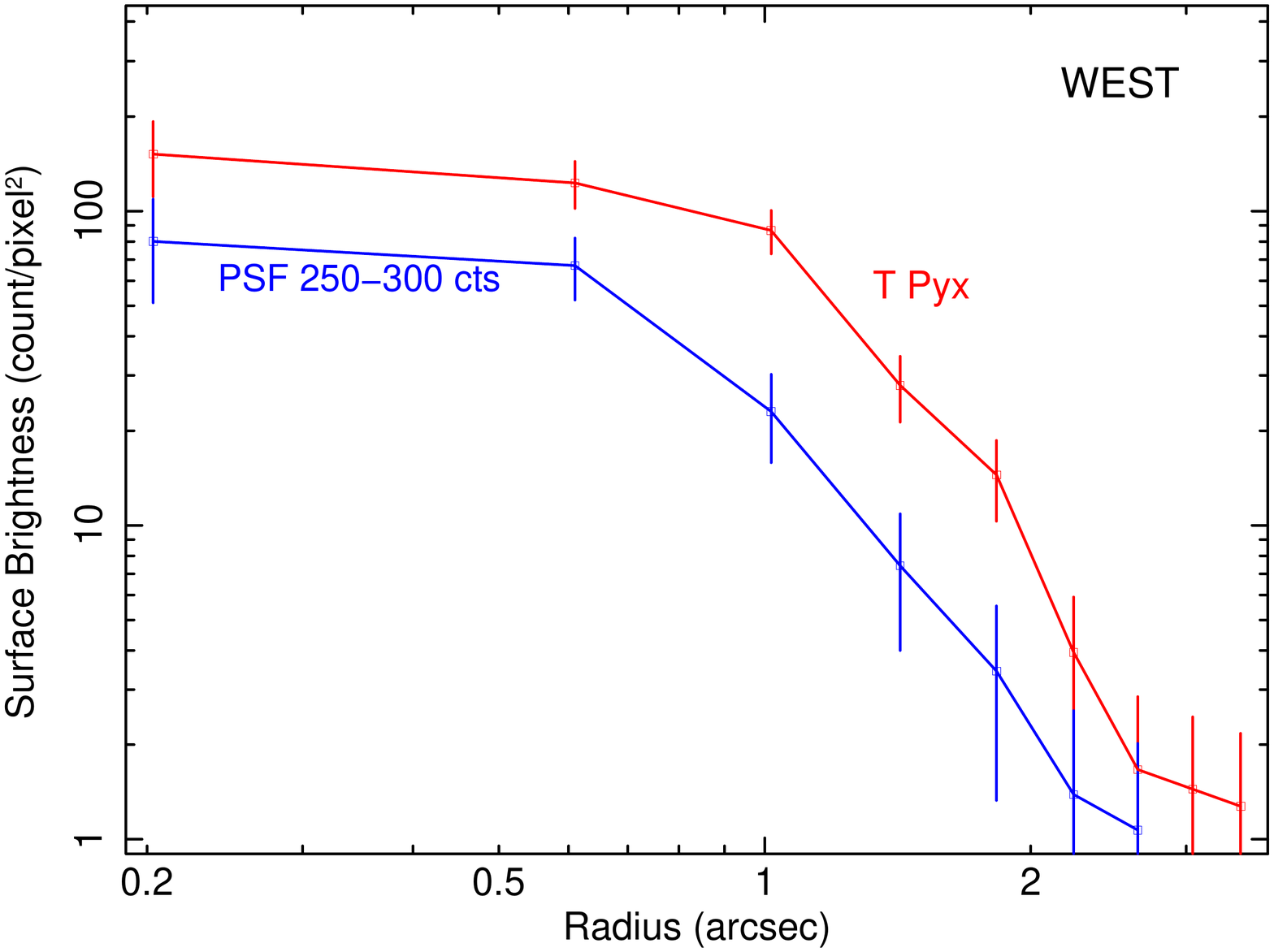}
\includegraphics[width=1.85in,height=1.95in,angle=0]{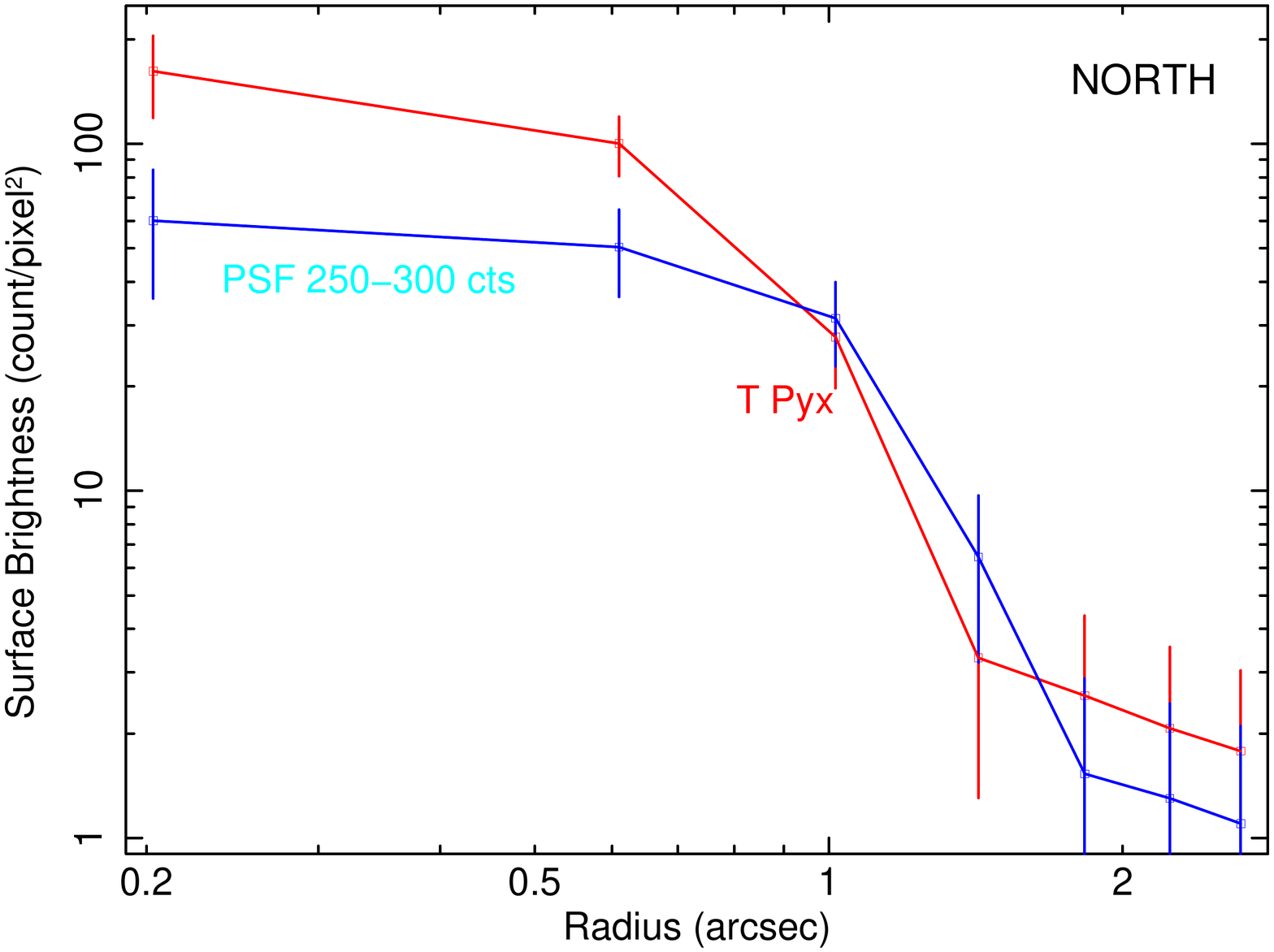}
\includegraphics[width=1.85in,height=1.95in,angle=0]{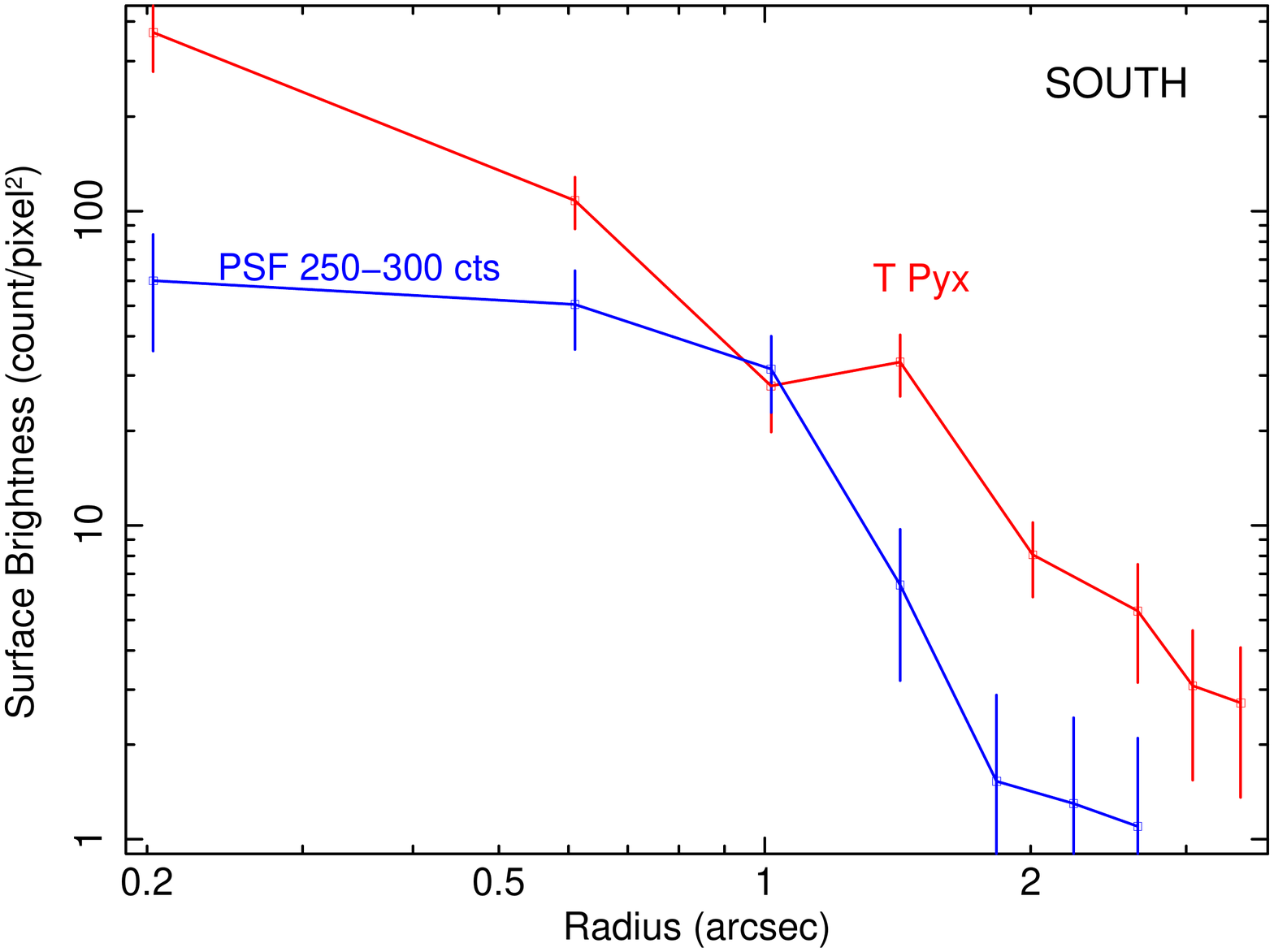}}
\caption{ The radial surface brightness profiles of the source T Pyx
extracted from a sector-area of opening angle 30 degrees.
From left hand to the right hand panels the radial profiles are in the
the eastern, western, northern and southern directions, respectively.
A \cha\ ACIS-S3 PSF radial profile of about 250-300 photons calculated from the
same sector is overplotted for comparison with Figure 4.
Gaussian errors are assumed.}
\end{figure*}

The deconvolved approximate nebular X-ray spectrum was modeled using a two-component plasma emission  model 
(in thermal equilibrium) with two
different hydrogen column densities since a single plasma model
or a single absorption component yielded \rchisq\ larger than 2 (see Figure 6 middle panel).
However, the second thermal
component and thus, the high absorption between two components
are found from a self-consistent analysis across all selected energy channels, but
need to be taken with caution as a result of the described caveats in the previous paragraph.
Table 2 shows the spectral
parameters for two-component plasma model fits with a MEKAL model
(assuming collisional equilibrium) and a PSHOCK model
(assuming non-equilibrium ionization). The fits yield similar parameters. The ionization timescales
derived from the PSHOCK model indicate existence of non-equilibrium ionization plasma,
but the parameter can not be constrained allowing for ionization equilibrium.
The spectral results (e.g., double MEKAL) indicate an absorbed X-ray
flux of (0.3-10.0)$\times 10^{-14}$ erg s$^{-1}$ cm$^{-2}$ which translates to
an X-ray luminosity of (0.6-30.0)$\times 10^{31}$ erg s$^{-1}$ at the source distance of 3.5 kpc. 
The count rate error of 40\% in the subtraction of the nebular
component is quadratically added to the normalization, flux and luminosity. 

\begin{table}[ht]
\label{2}
\caption{ Spectral Parameters of the Nebular Spectrum of T Pyx (0.2-9.0 keV);
ranges correspond to 2$\sigma$\ errors ($\Delta$\chisq = 3.84 -- single parameter)
for the first and 3$\sigma$\ errors ($\Delta$\chisq = 6.63 -- single parameter) for the second component;
$\chi^2_{\nu}$\ values of the fits are $\le$ 1.0.}
\begin{center}
\begin{tabular}{c|c|c} \hline\hline
\multicolumn{1}{c}{  } &
\multicolumn{1}{c}{\ PSHOCK$^{\S{1}}$} &
\multicolumn{1}{c}{\ MEKAL$^{\S{2}}$} \\
\hline

 \nh$_1$ ($\times 10^{22}$ cm$^{-2}$) &  0.4$^{+0.25}_{-0.2}$ &
0.5$^{+0.4}_{-0.3}$  \\

kT$_{s1}$ (keV) &  1.0$^{+1.1}_{-0.4}$ &
0.6$^{+0.3}_{-0.3}$ \\

${\tau}_1$ ($\tau$=n$_0$t) ($\times 10^{11}$ s cm$^{-3}$) & $\sim$3.2 & N/A  \\

K$_1$$^{\S{3}}$\  ($\times 10^{-5}$ cm$^{-5}$) &
0.6$^{+0.3}_{-0.5}$ & 0.6$^{+0.3}_{-0.3}$  \\

\hline
\hline

 \nh$_2$ ($\times 10^{22}$ cm$^{-2}$) &  11.5$^{+22.5}_{-7.3}$ &
 17.0$^{+9.7}_{-14.0}$  \\

kT$_{s2}$ (keV) &  2.3$<$ &
2.2$^{+2.0}_{-1.2}$   \\

${\tau}_2$ ($\tau$=n$_0$t) ($\times 10^{11}$ s cm$^{-3}$) & $\sim$2 & N/A  \\

K$_2$$^{\S{3}}$\  ($\times 10^{-5}$ cm$^{-5}$) &
3.5$^{+5.2}_{-1.7}$ & 9.8$^{+23.0}_{-7.5}$  \\

\hline
\end{tabular}
\end{center}
{\bf Notes.}
{\bf \S{1}} {The  model is ($tbabs$*PSHOCK+$tbabs$*PSHOCK);
$tbabs$--Wilms et al. 2000; PSHOCK--Borkowski et al. 2001.}\ 
{\bf \S{2}} {The  model is ($tbabs$*MEKAL+$tbabs$*MEKAL);
MEKAL--Mewe et al. 1986.}\ 
{\bf \S{3}} {Fit normalizations; a propagated count rate error of 40\% is quadratically added.
The normalization constant of the MEKAL/PSHOCK plasma emission models is
K=(10$^{-14}$/4$\pi$D$^2$)$\times$EM where EM (Emission Measure) =$\int n_e\ n_H\ dV$
(integration is over the emitting volume V).}
\end{table}

To determine the approximate central source spectrum (the binary), I fitted the combined spectrum with a
composite model
($tbabs$*(MEKAL+CEVMKL)+$tbabs$*MEKAL).
The two MEKAL  models (with the two different
absorption models) represent the nebular spectrum and the  CEVMKL is used to model the
central binary spectrum typical of accreeting nonmagnetic CVs.
 The nebular spectrum parameters are slightly varied
around their best fit values (the lower plasma temperature of the nebula and the \nh$_2$ is slightly varied)
during the fitting procedure. The fitted combined spectrum is
displayed in the right hand panel of Figure 6.
Using the CEVMKL model, the spectral parameters of the central source are a kT$_{max}$ $>$ 14.0 keV,
and a normalization of 5.1$^{+1.6}_{-1.0}$$\times 10^{-5}$ for the
spectrum of the central source emission (the power law index of $\alpha$ for the temperature
distribution was fixed to 1.0). This yields an unbsorbed X-ray flux of (2.0-7.0)$\times 10^{-14}$
erg s$^{-1}$ cm$^{-2}$ with a luminosity of (2.9-10.3)$\times$10$^{31}$ erg s$^{-1}$ (0.2-9.0 keV)
(40\% error in the subtraction of the nebula is quadratically added
to the normalization, flux and luminosity). 
The \nh\ parameter
is the same as in Table 1. The \rchisq\ of the fit with the CEVMKL model
for the central source spectrum is 0.6.
For the sake of the caveats mentioned on the second component of the nebular emission, 
I have assumed only the first low temperature component of the nebular
spectrum omitting the second component and fitted the combined spectrum to check the central source
spectral parameters. The CEVMKL fit yields kT$_{max}$$>$ 39 keV ($\alpha$ fixed at 1.0) 
with a similar range of interstellar
\nh\ together with flux and luminosity in a similar range to the analysis in section (3.1) 
(given inclusion of the 40\% error in the subtraction of the nebula). 

\begin{figure*}
\centerline{
\includegraphics[width=2.7in,height=2.6in,angle=0]{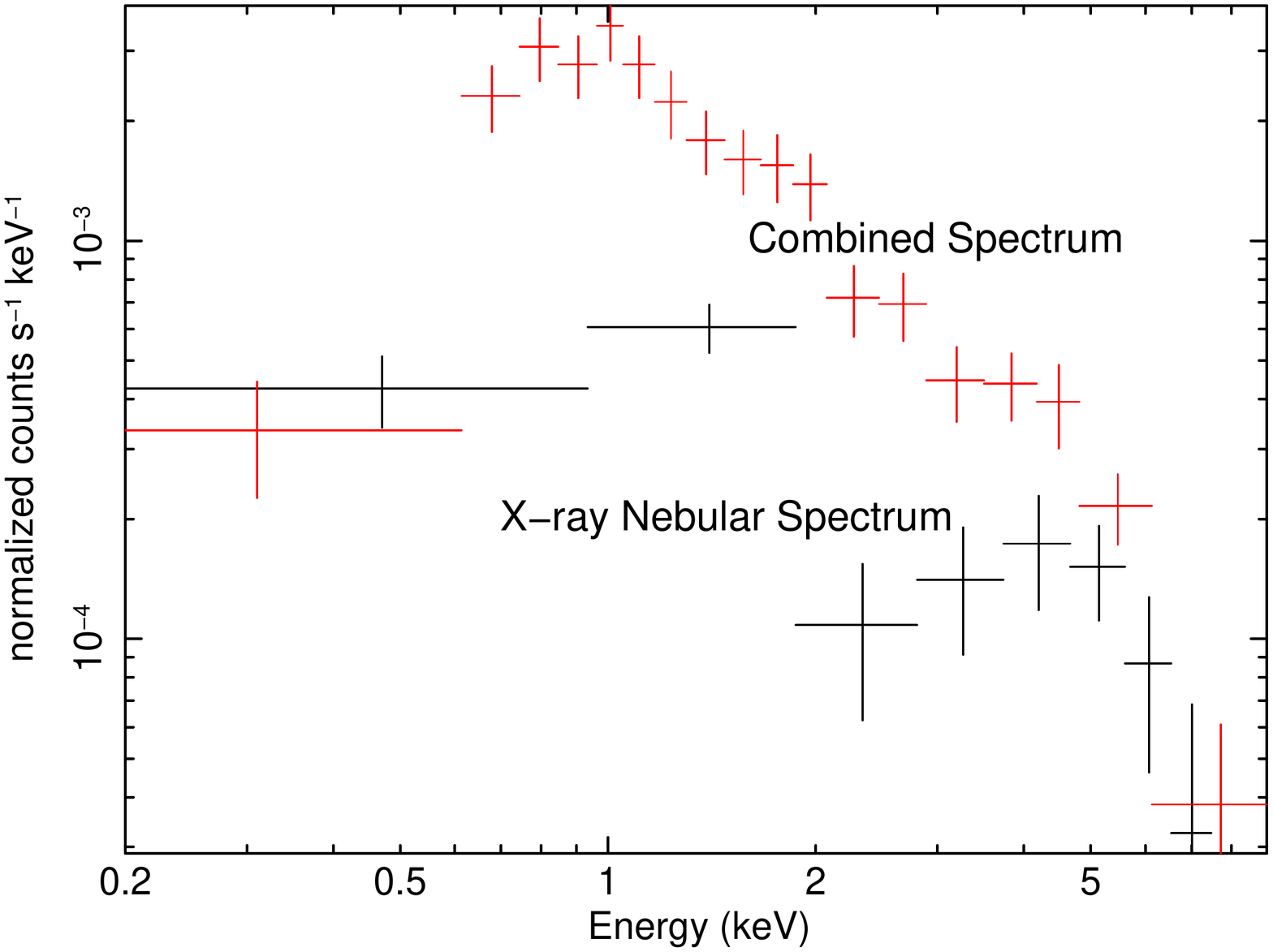}
\hspace{-0.55cm}
\includegraphics[width=2.7in,height=2.6in,angle=0]{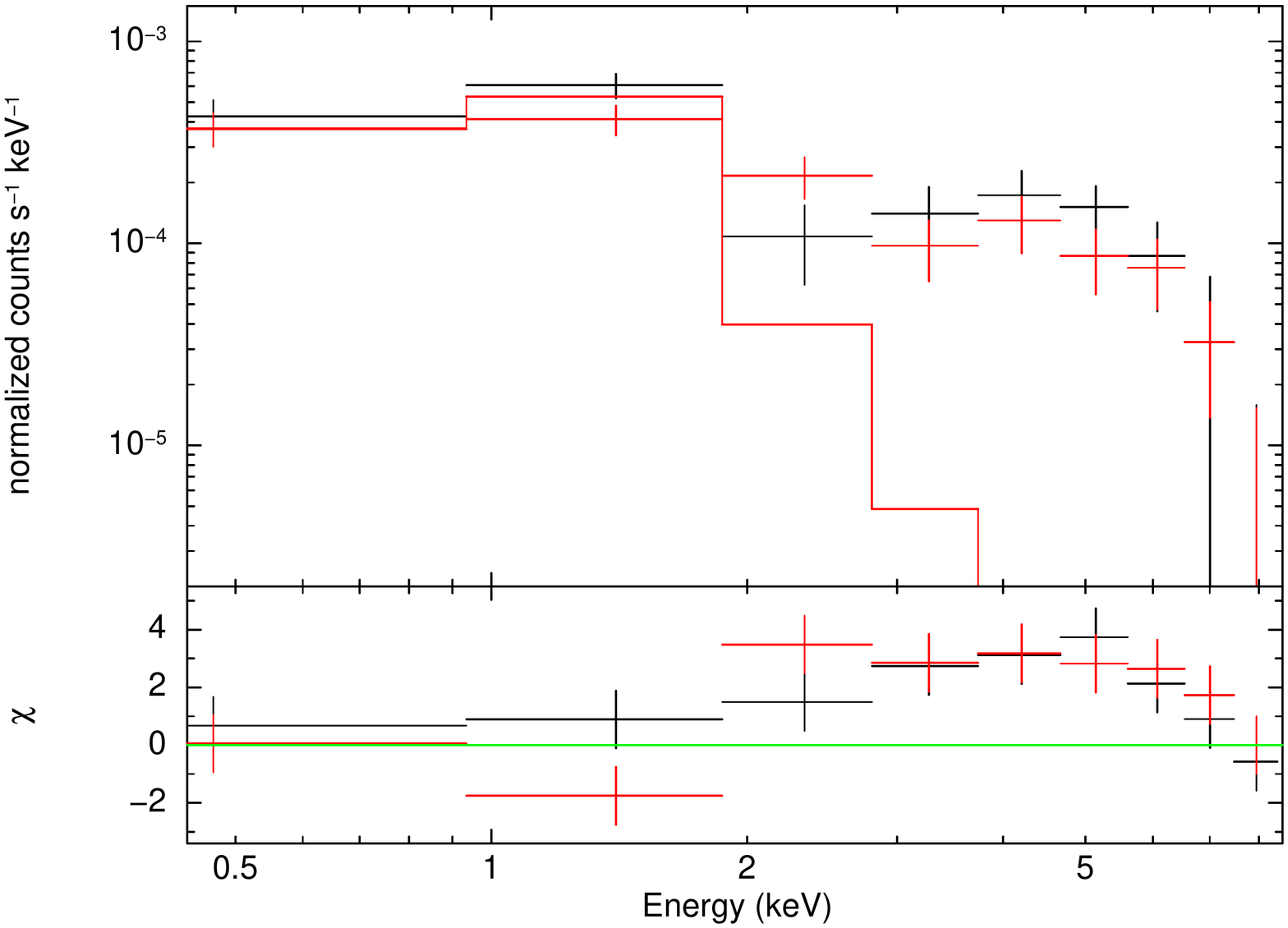}
\hspace{-0.55cm}
\includegraphics[width=2.7in,height=2.6in,angle=0]{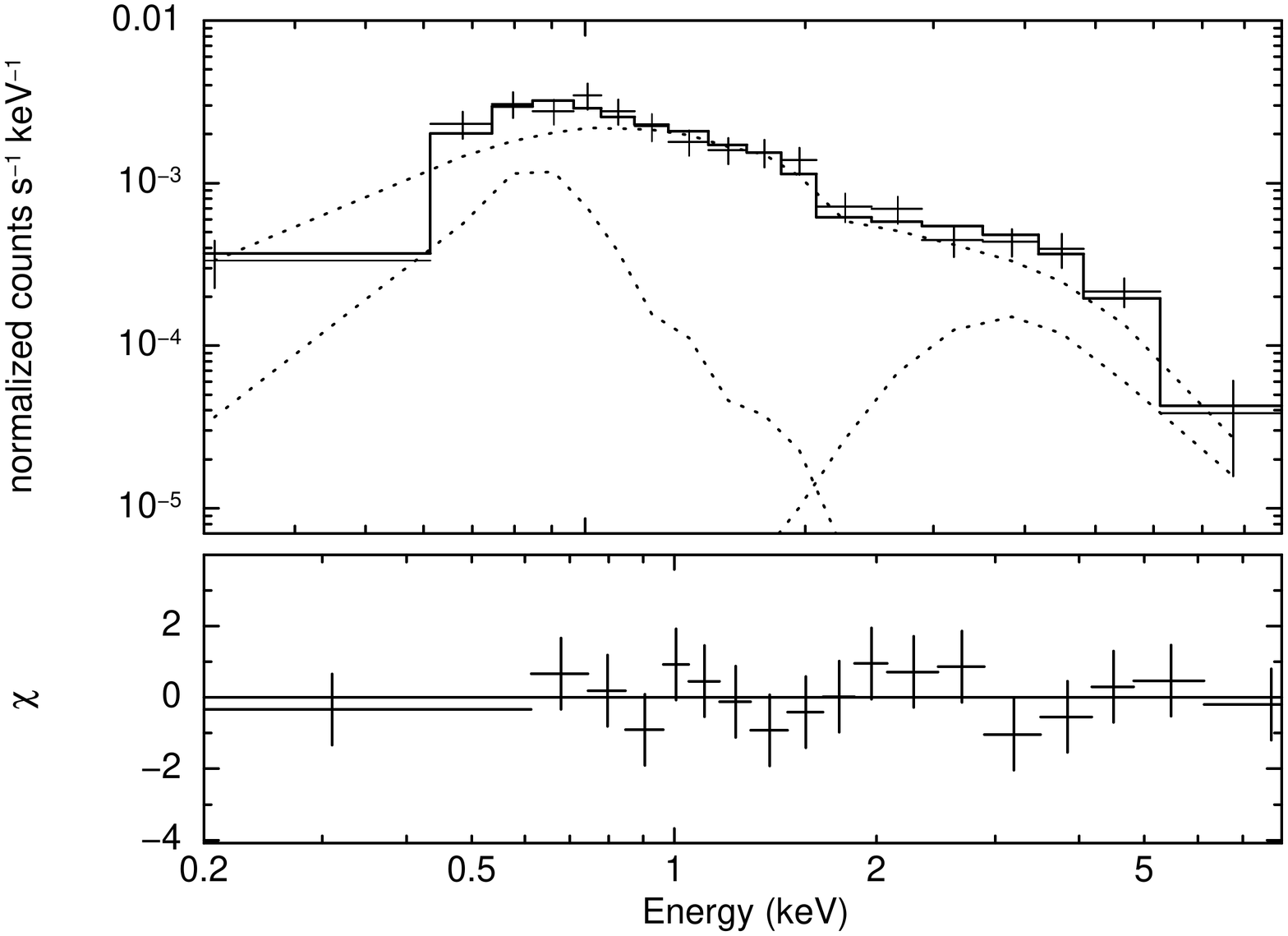}}
\caption{The left hand panel shows the combined/total spectrum of the source plus the nebular emission in the
X-rays (in red) and the nebular X-ray spectrum of T Pyx (in black). The middle panel
demonstrates a fit of a single componant plasma model to the X-ray nebular spectrum and the residuals in sigmas.
Black denotes the  spectrum derived from the 2$\times$2 pixel smoothed images (also on the left) 
and the red spectrum is the one derived from the 3$\times$3 pixel smoothed images. 
The right hand panel
shows the Chandra ACIS-S3 combined spectrum fitted with ($tbabs$*(MEKAL+CEVMKL)+$tbabs$*MEKAL) model of
emission.  The dotted lines show the contribution of the three fitted models two of which fits the X-ray
nebular spectrum and the third fits the central source spectrum.
The lower panel shows the residuals  in standard deviations (in sigma).}
\label{fig:spect}
\end{figure*}

\section{Discussion}
\label{sect:discussion}

\subsection{The Central Source/Binary}

I have presented the pre-outburst spectral and temporal analysis of T Pyx in quiescence
obtained around early Feb 2011 in three 30-ksec observations within about a week.
The total spectrum of the combined three datasets yields
hard X-ray plasma emission from the source with an X-ray temperature of
kT$_{max}$$>$47.0 keV (a 2$\sigma$ lower limit from the
CEVMKL fit is 37 keV), an \nh\ of 0.08$^{+0.10}_{-0.05}$$\times 10^{22}$ cm$^{-2}$ and
an unabsorbed X-ray flux of (5.0-9.0)$\times 10^{-14}$  \flux\  with a range of X-ray luminosity
(0.8-1.4)$\times 10^{32}$ \ergsec\ in 0.2-9.0 keV band. 
In a wider range of 0.1-50 keV the integrated flux and luminosity
are (0.9-1.5)$\times 10^{-13}$  \flux\ and
(1.3-2.2)$\times 10^{32}$ \ergsec, respectively.
A multi-temperature plasma emission model, CEVMKL, is assumed for the fit in accordance 
with modeling of the nonmagnetic CVs in quiescence.
A double MEKAL model is also consistent with the total X-ray spectrum, reflecting that the plasma temperature is
already a multiple temperature distribution.
On the other hand, the lower plasma temperature resembles the temperature of the
extended nebulosity recovered from the source in this work and also using $XMM$-$Newton$ 
which, then, may belong to a different origin. I note that the combined spectrum show 
consistency with Comptonized plasma emission models, and they reveal a second thermal
plasma component with a temperature in accordance with the low temperature of the extended nebulosity.   

The optical V band magnitude of T Pyx was in a range 15.4-15.6  in 2006
during the $XMM$-$Newton$ and in 2011 during  the \cha\ observations
(using available AAVSO data).
Thus, the system brightness has not changed much in the optical band, which otherwise might have
signaled a noticeable change in the accretion state. The comparison of the total X-ray flux (unabsorbed)
and the X-ray luminosity of the central source in T Pyx (in the 0.2-10.0 keV range) indicates that
there is definite overlap of the flux ranges calculated for the
source in  2006 ($XMM$-$Newton$) and 2011 (\cha) observations
since the error on the unabsorbed flux in  $XMM$-$Newton$ data is large and between
(3.0-80.0)$\times 10^{-14}$  \flux.

A complicated deconvolution of the central source spectrum (see sections [3.3] and [3.4])
has been applied to the data to extract
any extended X-ray nebulosity from the central source emission, the results indicate that
the X-ray temperature of the central source is still hot
with kT$_{max}$$>$ 14 keV (best fit results using a MEKAL and CEVMKL model,
respectively). The maximum temperature of the plasma temperature distribution is unconstrained.  
The corrected unabsorbed flux and luminosity are 
(0.7-7.0)$\times 10^{-14}$
erg s$^{-1}$ cm$^{-2}$ and (1.1-10.0)$\times$10$^{31}$ erg s$^{-1}$, respectively. 
This luminosity and flux are (3.0-12.0)$\times 10^{-14}$ erg s$^{-1}$ cm$^{-2}$ and
(4.4-18.0)$\times$10$^{31}$ erg s$^{-1}$\ in the 0.1-50 keV energy range. 

I do not find a blackbody emission component consistent with (satisfactorily fitting with significance) 
the \cha\ pre-outburst spectra as was detected with
$XMM$-$Newton$ (Selvelli et al. 2008, Balman 2010) and $ROSAT$ 
(RASS data; Greiner \& Di Stefano 2002) in the earlier studies
over the years (there has been no Super Soft X-ray source associated with T Pyx in quiescence).
As mentioned in section (3.1), there is no intrinsic absorption in the system, but absorption only at 
the interstellar value. 
I exploit the data to calculate a 2$\sigma$\ upper limit to temperature and flux for
the blackbody emission from T Pyx as kT$_{BB}$$<$25 eV, f$_{BB}$$<$1.5$\times$ 10$^{-12}$ \flux\ and
L$_{BB}$$<$2.0$\times$10$^{33}$ \ergsec\ in the
0.1-10.0 keV band. 

Gilmozzi \& Selvelli (2007) have studied the UV spectrum of T Pyx in detail and found that 
the spectral energy distribution (SED)
is dominated by an accretion disk in the UV+opt+IR ranges, with a distribution, described by a power law
$F_{\lambda}$ = 4.28$\times$10$^{-6}$ $\lambda$$^{-2.33}$ \flux\ \AA$^{-2}$, 
while the continuum in the UV range,
alone, can also be well
represented by a single blackbody of T $\sim$ 34000 K. The observed UV continuum distribution of T Pyx has
remained constant in both slope and intensity during 16 years of IUE observations.
Selvelli et al. (2008) predict a disk luminosity of about
3$\times$10$^{35}$ erg\ s$^{-1}$ (from UV and optical bands) for T Pyx consistent with the accretion
rate they calculate in the optical and UV rejimes (1-4)$\times$10$^{-8}$ M$_{\odot}$ yr$^{-1}$ 
for a WD mass of(0.7-1.4)\msol. I note that larger accretion rates of 
(1-10)$\times$10$^{-7}$ M$_{\odot}$ yr$^{-1}$ have been suggested for T Pyx, as well (Schaefer et al. 2013,
Godon et al. 2014).
In this accretion rate range, the nature of the boundary layer should be discussed in the framework of
the calculations of the standard steady-state
disk models expected for CVs, e.g. of Narayan \& Popham (1993), Popham \& Narayan (1995), Popham (1999).
These models predict optically thick BLs with blackbody
temperatures of 13-33 eV and L$_{soft}$$\ge$ 1$\times 10^{34}$ for 0.8-1.0 \msol\ WD (a 
rotation as high as $\Omega_{*}$=0.5$\Omega_K(R_{*})$ is already assumed in luminosity/temperature limits) 
which I did not recover using the \cha\ data.
I emphasize here that the standard nova theory predicts near chandrasekhar masses for the WDs in RNe
and this predicted (and observed in other RNe) mass value implies much hotter and X-ray luminous optically
thick boundary layer for T Pyx than these given range limits. 
 
Detailed calculations by Narayan \& Popham (1993) show that the optically thin BLs of accreting WDs
in CVs can be radially extended and that they advect part of the
internally dissipated energy as a consequence of their inability to cool, therefore
indicating that optically thin BLs act as ADAF-like accretion flows.
In addition, Popham \& Narayan (1995) illustrates that
the BL can stay optically thin even at high accretion rates (as for T Pyx) 
for optical depth $\tau$ $<$ 1 together with
$\alpha$\ $\ge$ 0.1\ . However, nature of such models are not well investigated.
An ADAF around a WD can be described by truncating the ADAF solutions at the WD surface
as opposed to BHs and the accretion energy is advected onto the WD heating it up.
Medvedev \& Menou (2002) and Menou (2000) include some preliminary work regarding
ADAF-like flows and hot settling flows in CVs (dwarf novae) where Menou (2000) suggests
that ADAF-like gas flows in the BLs of CVs are expected to be one-temperature in CIE since Coulomb 
interactions at low temperatures compared with BH binaries are efficient and also advection 
does not depend on preferential heating of the ions by viscous dissipation and lack of energy exchange
between electrons and ions necessary for two-temperature flows as in BHs (NEI, non-equilibrium
ionization). However, since there are no detailed calculation of ADAF-like flows for CVs in general
the two-temperature flows can not be completely ruled out.
Recently, Balman et al. (2014) have shown that some Nova-like CVs have BLs that can be characterized
with ADAF-like flows merged with optically thin BLs 
in high state CVs. These objects have accretion rate similar or somewhat less
than T Pyx. The X-rays have optically thin multiple-temperature cooling flow type emission
spectra with temperatures kT$_{max}$ in a range 21-50 keV. These BL regions are also found 
divergent from the isobaric cooling flow models and the temperatures are at/around virial values
with similar characteristics to quiescent X-ray emission of T Pyx. 

Given the disk luminosity mentioned previously for T Pyx 
and the X-ray luminosity in the 0.1-50 keV range
the ratio (L$_{x}$/L$_{disk}$)$\simeq$(2-7)$\times$10$^{-4}$.
The nature of the boundary layer in the central source in T Pyx is not consistent with the predictions from
the calculations of the standard steady-state
disk models (L$_{x}$$\sim$L$_{disk}$), e.g. of Narayan \& Popham (1993), Popham \& Narayan (1995).
I note that discrepancy between L$_{x}$ and L$_{disk}$ is common for nonmagnetic CVs
and this ratio has been found to be around 0.1 for SU UMa type dwarf novae and about 0.01 for the U Gem
sub-type together with the nova-likes at high states 
having a ratio of around 0.001 (see Kuulkers et al. 2006 for a review). 
The expected soft X-ray/EUV emission is not detected from the BL region and 
the accretion flow is most likely virialized and very hot and optically thin as discussed in section (3.1).
I suggest that the X-rays from the point source originate in an optically thin boundary layer that
is merged with an ADAF-like flow
and/or in an accretion disk coronal region where accretion occurs through hot
coronal flows onto the WD. Such a quasi-spherical/torus-like
hot accretion flow in the BL zone may also obscure the WD even at low
inclination angles.

In X-ray binaries, particularly black hole systems,
an inner advection-dominated accretion flow (ADAF)
exists that extend from the black hole horizon to a
transition radius and above the disk there is a hot
corona which is a continuation of the inner ADAF
(Esin, McClintock, Narayan 1997; Narayan, Barret, McClintock 1997; Narayan \& McClintock 2008).
ADAFs are based on  $\alpha$-viscosity prescription where substantial fraction of the
viscously dissipated energy is stored in the gas and advected to the central object with the accretion flow
rather than being radiated. This may explain the orders of magnitude difference
in the X-ray luminosity and the accretion luminosity in the UV/optical bands for T Pyx.
If there is a corona, then the mass accretion rate
in the inner regions is maintained by the disk-corona interface rather than the secondary star.
Note that for dwarf novae the evaporation and disk truncation models by Meyer and Meyer-Hofmeister (1994)
and Liu et al. (1995) can enhance the accretion rates onto the WDs in the disk during the quiescent phases
by about 20 to 100 times 
(from eg. 1$\times 10^{-12}$ \msol\ yr$^{-1}$ to a few$\times 10^{-11}$ \msol\ yr$^{-1}$).

Some eclipse mapping studies of
quiescent dwarf novae indicate that mass accretion rate
diminishes by a factor of 10-100 and sometimes by 1000 in the inner regions
of the accretion disks as revealed by the brightness temperature calculations
which do not find the expected R$^{-3/4}$ radial dependence of brightness temperature
in standard steady-state
constant mass accretion rate disks (see e.g., Z Cha and OY Car: Wood 1990,
V2051 OPh: Baptista \& Bortoletto 2004, V4140 Sgr: Borges \& Baptista 2005).
Biro (2000) finds that this flattening in the brightness
temperature profiles may be lifted by introducing optically thick disk truncation in the quiescent state
which may lead to formation of hot accretion flows in the disk even in quiescent systems
(e.g., r $\sim$ 0.15R$_{L1}$ $\sim$4$\times$10$^{9}$ cm; DW UMa, a nova-like).
Balman \& Revnivtsev (2012), also finds optically thick disk truncation in a range of radii
(3.0-10.0)$\times$10$^{9}$ cm in quiescent dwarf novae with possible formation of hot flows 
(possibly ADAF-like) in the disk 
(see also Liu et al. 2008).
A comprehensive UV modeling of accretion disks
at high accretion rates (similar with T Pyx) in 33 CVs including many nova-likes and old novae
indicates departures from standard disk model (Puebla et al. 2007) with
an extra component from an extended optically
thin region (e.g., wind, corona/chromosphere) with NLTE effects and 
strong emission lines and P Cygni profiles observed in the
UV spectra. 

To explain the extremely blue color of T Pyx in quiescence (in the optical),
Webbink et al. (1987) proposed that nuclear burning
continues even during its quiescent state, consistent with the slow outburst development, 
which suggested that the accreted envelope was only weakly degenerate at the onset of TNR. 
Patterson et al. (1998) and Knigge et al. (2000)
attributed the luminosity of T Pyx to quasi-steady thermonuclear burning as a wind-driven 
supersoft X-ray source (SSS). However, T Pyx has never been detected as an SSS,
as mentioned earlier in this section, and thus the true accretion rate is expected to be lower than
what is assumed for SSS (i.e. $\dot{\rm M}$ $<$ 1$\times$10$^{-7}$). 

In general, bare accretion disks that produce substantial numbers of
ionizing photons have optical/UV SEDs that are too blue (e.g., AGN disks).
Schaefer et al. (2013) argues that the ultraviolet portion of the T Pyx SED is
consistent with a   $f\ \propto$ $\nu$$^{1/3}$ power law, but the entire UBVRIJHK part
of the SED has $f\ \propto$ $\nu$$^{1}$ which shows that the UBVRIJHK
emission is not from an accretion disk. Studies have shown that median optical to UV
continuum slope in radio-quiet QSOs/AGNs is also $f\ \propto$ $\nu$$^{1/3}$ power law
(Francis et al. 1991). One way of redistributing the energy and changing the IR-optical-UV continuum slope
is to assume that part of the optical/UV arises from reprocessing of the radiation from the inner disk,
a disk irradiated by a central source or hard X-ray emitting boundary layer
with flaring and/or warping further out in the disk, the changes in the slope
may be even more pronounced. Note that
such structures exist in AGN disks (see a review by Koratkar \& Blaes 1999).

Accretion disks may not be flat in shape if the angular momentum of the flow is misaligned
with the spin axis, then the disk will be warped e.g., in an AGN disk (Bardeen \& Petterson 1975).
It may also be that the warping is radiation-induced.
A warped disk intercepts and reprocesses radiation from the inner
regions, the effective temperature distribution could flatten
from the canonical r$^{-3/4}$ to r$^{-1/2}$ , which would produce
a long-wavelength SED of $f\ \propto$ $\nu$$^{1}$, much redder than the canonical
$\nu$$^{1/3}$ distribution. This may be what is observed by
Schaefer et al. (2013). Therefore, for understanding T Pyx,
the possibility of reprocessing of X-ray photons from the inner
disk (e.g., ADAF-like bounday layer) or X-ray corona by an outer warped, flared, or ruffled disk needs
to be explored similar to AGN-type disks, e.g., non-standard disks.

The orbital  modulations detected in T Pyx are characteristic of
magnetic or nonmagnetic CVs and can be caused by  scattering
of the X-rays from structures fixed in the orbital plane particulary if there are local regions
of high vertical extent (e.g., elevated disk rim or scattering from a coronal region).
A warped disk effect should also
be considered for these orbital modulations. A disk overflow (from the accretion impact region) 
may also modulate the boundary layer emission on the binary period. 
Note that the orbital variations are superimposed
over an unchanging constant component of 0.003 c s$^{-1}$ which indicates that there is either
a scattered isotropic unvarying componant in the central source emission or it is a signature of the
extended emission component that does not have a direct connection to the binary system
(see Figure 2 and section [3.2]).

\subsection{An Evaluation on the X-ray Emitting Nebula}

An analysis of the $XMM$-$Newton$ EPIC pn data of T Pyx showed deviations in the surface brightness
(radial) profiles
from a standard source PSF (Balman 2010). The
low pixel resolution (4\arcsec ) and the large PSF size
(PSF core radius $\sim$  6\arcsec-7\arcsec; Str\"uder et al. 2001)
hindered the spatial deconvolution.
An extended emission around 7\arcsec-15\arcsec\
was found consistent with data as the fits using model PSF yielded large \rchisq\ values
between 2\arcsec-20\arcsec\ . On the other hand, this result
is constrained with the half energy width size (HEW around 15\arcsec, kernel size/radius of the PSF) of
the HRMA+EPIC pn PSF. This limits
the maximum size at about 15\arcsec\ for the X-ray nebular emission without a limit on
the smallest size for any weak emission. Note that the radius of the  optical remnant is
about 5\arcsec-6\arcsec which falls in the PSF core region.
I stress that there is no other known detected X-ray source within 20 arcsec
of T Pyx. In addition, there are no nearby sources detected with the
sensetive $XMM$-$Newton$ Serendipitious Survey/Catalogue
within this range. There is a source at (09:04:43.5, -32:22:60) location close to T Pyx,
this source is 32\arcsec.5 away from T Pyx which is twice the HEW of $XMM$-$Newton$ EPIC pn.
Therefore, this source is resolved in the spatial analysis 
and does not effect the fits performed on the radial profiles.
The  $XMM$-$Newton$  EPIC pn detection is
at about 4-5$\sigma$ level (calculation of $\sigma$ level explained in sec. 3.1)
and the count rate ratio between XMM MOS and pn detectors is about 1:3 (for the given observing
modes), respectively. This translates to about 2-2.3$\sigma$ excess detection with the
MOS detectors. Given the low statistical quality with lower count rates from the MOS detectors,
errorbars of the radial profiles are larger  
yielding  \rchisq\ values around 1.0 or less for the fits.

There is an archival observation of T Pyx obtained with \cha\ LETG/HRC-S
in November 3, 2011 during the outburst stage. The average count
rate of the X-ray nebula around 0.0025 c s$^{-1}$ translates to a rate of 0.00015 c s$^{-1}$
(using PIMMS v4.6b, WebPIMMS-http://heasarc.nasa.gov/Tools/w3pimms.html)\ for the
X-ray nebula with the LETG/HRC-S observation yielding only six nebular photons
in the 39.8 ksec observation with no significant detection. T Pyx is in outburst so the source is
very bright in the zeroth order.

There is another archival $XMM$-$Newton$ observation of T Pyx obtained on November 28, 2011
during the outburst. The EPIC pn
net source rate is about 1.2 c s$^{-1}$  (21.3 ksec exposure) and the EPIC MOS net source rate is about
0.25 c s$^{-1}$  (30.5 ksec exposure). The expected X-ray nebular source
photons of about 16-20 (in MOS) and
40-46 (in pn), yield a sigma detection less than 0.1,
therefore data can not be used to derive the nebulosity.

The $XMM$-$Newton$ detection and the \cha\ ACIS-S3 nebulosity is, in general, consistent. 
The excessive elliptical nebula has size about 0.\arcsec5-0.\arcsec9 
outside the ACIS-S3 PSF core (0.\arcsec42 PSF core size, and HEW about 0.8-1 arcsec).
The $XMM$-$Newton$  PSF core
is about 6\arcsec\ - 7\arcsec\ (EPIC pn) with a spatial resolution of 4\arcsec\ and the HEW (half
energy width) or the PSF kernel size/radius of about 15\arcsec. Please note carefully the
capability differences of EPIC pn and Chandra ACIS-S3.
The EPIC pn analysis (Balman 2010) shows
distorted PSFs with some excess within about a region of 15\arcsec.
Therefore ACIS-S and EPIC pn show excessive emission around and/or larger than
the core radius within the size of the PSF kernel, meaning inside the PSF region.
Note that at 4\arcsec\ spatial resolution, $XMM$-$Newton$ image of T Pyx is an ellipsoid.
In general, one can state that $XMM$-$Newton$ EPIC pn detects, but does not
resolve the nebula whereas \cha\ ACIS-S3  detects and resolves the nebula with
the high-resolution imaging capability at the sub-pixel level.

The detailed  high-resolution imaging deconvolution of T Pyx at the sub-pixel level 
indicates the possible
existence of a elliptical structure (possibly torus or ring-like)
around T Pyx which most likely suffers projection effects (e.g. east-west contrast, see 
section 3.3).
If one assumes a simple circular region viewed along a given line of sight (inclination) angle
(R sin(i)= semi-minor axis) taking R=0\arcsec.9 and semi-minor axis as 0\arcsec.4, one finds
an inclination angle of $i \le 27^{\circ}$. If one assumes results from the calculations
of surface brightness profiles from 30 degree-wide sectors, taking R$\simeq$1\arcsec.5
and semi-minor axis $\simeq$0\arcsec.5, the inclination angle is $i \le 20^{\circ}$.
The inclination of the binary system is about 10$^{\circ}$ (Uthas et al. 2010).
The elliptical structure may be an ejected interacting shell or may be a projection effect of a
conical face-on bipolar ejection interacting with a large spherical shell,
where the cut plane will be in the form of an ellipse. The elongation towards south
may also be related to a bipolar ejection from the nova in one of the outbursts.
A face-on bipolar ejection was detected in the 2011 outburst of T Pyx using the near-infrared
observations (Chesneau et al. 2011). The small size of the X-ray nebula indicates that it most 
likely results from an
interaction of the 1966 ejecta with the pre-existing shells/older ejecta.
The size of the torus-like structure $\sim$ 0\arcsec.9 (a possible minimum due to projection effects)
yields an expansion velocity of V$_{exp}$$\ge$400D$_{3.5kpc}$ km s$^{-1}$.

The spectral deconvolution shows at least one plasma emission component
from the X-ray Nebula.
The calculated emission measure EM=$<n_e>^2$ V$_{eff}$, obtained from the normalization of the fit,
yields an average electron density n$_e$ of about 32 cm$^{-3}$ 
for the colder plasma (a spherical region of 1\arcsec\  radius at 3.5 kpc
is assumed with a filling factor f=1).
The shocked mass in the X-ray nebula of T Pyx can be approximated as
$\le$1.8$\times 10^{-5}$M$_{\odot}$
assuming a fully ionized gas, and
M$_{neb}$ $\simeq$ n$_e$ m$_H$ V$_{eff}$. By comparison, Selvelli et al.
(2008) have calculated an ejecta mass of 10$^{-5}$-10$^{-4}$M$_{\odot}$ for the 1966 outburst.
Shore et al. (2011) measured an ejecta mass
of $<$1$\times 10^{-5}$M$_{\odot}$ (f=1) for the 2011 outburst. Patterson et al. (2013) have
calculated that the total ejecta mass is
around 3$\times 10^{-5}$M$_{\odot}$ for the same new outburst. Assuming that about 10\%
of the ejecta will be shock heated
the calculated shocked mass and the ejecta mass are in accordance.
The plasma temperature derived from the fits can be used to approximate
the shock velocities using the strong shock relation
$kT_s=(3/16)\mu m_H (v_s)^2$, (T = 1.4$\times$10$^5$ v$^2_{100 km s^{-1}}$).
I derive about 500-870 km s$^{-1}$ for the first component consistent with
the minimum speed  calculated from the size of the X-ray nebula in the previous paragraph
and the expansion speeds of the 1966 outburst that are in the range
850-2000 km s$^{-1}$ (Catchpole 1969).

I speculate that the first plasma component is the forward shock and the second embedded
component is possibly due to the reverse shock. However, I caution that
there may be other speculative components of extended emission from T Pyx, not relating to the
nova remnant e.g., episodic/occasional jet outflows (see Shahbaz et al. 1997).
Contini \& Prialnik (1997) have modeled the circumstellar interaction of the T Pyx shells
where a forward shock moves into the older
ejecta and a reverse shock moves into the new ejecta.
The model predicts a
faster, hotter and denser reverse shock than the forward shock. \cha\ results are in  reasonable
agreement with the predictions in this paper.

\section{Summary and Conclusions} 
\label{sect:conclusions}

This paper presents the analysis of the pointed set of $\sim$ 3$\times$30 ksec \cha\ ACIS-S3 observations
of the recurrent nova T Pyx obtained in 2011 before its outburst in April 2011.
The total source spectrum is consistent with a multi-temperature (distribution) thermal plasma
emission  model (e.g. CEVMKL) (rather than a single temperature thermal plasma (e.g. MEKAL)
or a single power law model) with a maximum temperature
kT$_{max}$ $>$ 47.0 keV (2$\sigma$ lower limit is 37 keV), and
an unabsorbed X-ray flux of (0.9-1.5)$\times 10^{-13}$  \flux\  with an X-ray luminosity
(1.3-2.2)$\times 10^{32}$ \ergsec\ in the 0.1-50.0 keV.
The maximum plasma temperatures are virialized and thus, the plasma is likely not confined to the disk.
The absorption towards the source is at the interstellar level
consistent with the E(B-V) values as determined from the X-ray fits. The combined X-ray spectrum of T Pyx
shows consistency with Comptonized plasma emission models, particularly including an additonal 
thermal CIE plasma model with temperatures 0.2-0.55 keV.
The standard disk theory at steady state with constant $\dot{\rm M}$ predicts,
an optically thick boundary layer at the accretion rate of T Pyx 
($\dot{\rm M}$ $\ge$ $\times$10$^{-8}$ \msol\ yr$^{-1}$),
with a blackbody emission from the boundary layer in the soft X-ray/EUV regime.
I find no such emission with a  2$\sigma$\ upper limit to the blackbody temperature
kT$_{BB}$$<$25 eV and unabsorbed flux $<$1.5$\times$ 10$^{-12}$ \flux\ at
L$_{BB}$$<$2.0$\times$10$^{33}$ \ergsec\ (0.1-10.0 keV). This result is consistent with
all such results attained from the X-ray wavelengths. 
I underline that these upper limits on temperature and flux is particularly not in accordance
with the deep gravitational potential well of a near-chandrasekhar WD for production of
boundary layer emission from a rotating WD as high as $\Omega_{*}$=0.5$\Omega_K(R_{*})$.  
Moreover, these upper limits does not confirm any
SSS emission from a suggested H-burning WD during quiescence and constrains the maximum accretion rate
that can be attained ($\dot{\rm M}$ $<$ 1$\times$10$^{-7}$ \msol\ yr$^{-1}$).
In addition, I note that T Pyx does not have detected wind emission (see Uthas et al. 2010).
However, these upper limits are not so low to constrain any blackbody emission, for instance, 
that may originate from the disk itself, particularly in the inner disk where the 
UV emission is produced.

The X-ray to disk luminosity ratio of T Pyx is  (L$_{x}$/L$_{disk}$)$\simeq$(2-7)$\times$10$^{-4}$
given an L$_{disk}$ of 3$\times$10$^{35}$ \ergsec\ and the ratio will be even lower for any
L$_{disk}$ larger than this.
These results strongly indicate that the boundary layer of T Pyx is non-standard.
I suggest that the X-ray emission is due to an optically thin
boundary layer merged with an ADAF-like hot flow and/or X-ray corona in the inner disk
where the flow does not radiate much and the accretion energy is advected onto the WD in likely a 
quasi-spherical geometry or torus-like geometry. The advective heating of a
lower mass WD (0.7-1.0 \msol\ ) may possibly yield the onset of a TNR (thermonuclear runaway)
together with the (very) high accretion rate in the system causing the RN event.
However, X-rays do not constrain the mass of the WD in the system for this new boundary layer
scenario to hold. I underline that Standard nova theory requires near-Chandrasekhar mass WDs
with high accretion rates (1$\times$10$^{-(8-7)}$ \msol\ yr$^{-1}$) for Recurrent novae. 
TNR and the recurrence of the TNRs are strong functions of the WD 
temperature and luminosity as much as
the quiescent accretion rate, WD mass, accreted mass, and metal abundances.
Note that an advected luminosity equivalent of a disk luminosity of 3$\times$10$^{35}$ \ergsec\ 
will yield a WD effective temperature of 2$\times$10$^{5}$ K
for a 1 \msol\ WD, consistent with effective WD temperatures during the onset of TNRs 
(see Starrfield et al. 2000).  I note that a  sub-Chandrasekhar mass WD in T Pyx may explain the
large ejecta mass detected in all RN explosions of this source together with the evolution
of the T Pyx light curves that are more typically found in slow/er novae associated with lower mass WDs
without the need to invoke delayed ejection scenarios as in Chomiuk et al. (2014).
However, I emphasize that this is an uncharted area in the 
field of novae and thus new theoretical calculations on this issue will help to improve our 
understanding.

In conjunction with the IR-optical-UV SED of the source/disk that shows a non-standard accretion
disk as a result of the power law dependence (aside from the UV), 
the possibility of reprocessing of photons from the inner
ADAF-like flow or an X-ray corona by an outer possibly warped, flared, a ruffled disk needs
to be considered for T Pyx in possible relation to AGN disks that has been observed to show similar 
type of power law dependence of SED. 

The binary period of the system is detected in the X-ray wavelengths with 
no energy dependence (over the orbit) in the
0.2-9.0 keV \cha\ band. The cause of the modulations may be the scattering from structures
fixed on the orbital plane, e.g., on the disk with high verticle extend. It is possible that
disk overflow with/or a warped nature of the disk and stream impact on the X-ray emitting region
may result in such orbital variations.

A detailed high-resolution imaging analysis at the sub-pixel level 
indicates excess emission around T Pyx with S/N $\sim$(6-10) 
as an elliptical nebulosity (possibly ring-like) with a
semi-major axis $\sim$ 1 arcsec and semi-minor axis $\sim$ 0.5 arcsec.
There is some elongation towards south ( $\sim$ 1.85 arc sec).
The north-south inclination angle of the elliptical nebula is  $i \le 27^{\circ}$.
Note that the radial profiles of the
smaller 30 degree sector show some evidence that the elliptical nebula  may be inclined also in
the east-west direction where west and south sides are towards and the
east and north sides are away from the observer.
Given the characteristics and the size of the elliptical nebulosity,
it seems consistent with an interaction of the ejecta
from the outburst in 1966. The extended emission excess is
within  2\arcsec-2.\arcsec5 of the central source and
shows emission from plasma in ionization equibrium or close to equilibrium
with plausibly two temperatures around 0.6 keV and 2.2 keV. Note that the
second harder X-ray temperature component has caveats from the spectral deconvolution process. 
The nebular flux and luminosity are  (0.3-10.0)$\times 10^{-14}$ erg s$^{-1}$ cm$^{-2}$ and
(0.6-30)$\times 10^{31}$ erg s$^{-1}$ (at 3.5 kpc) with considerable uncertainty. I note that 
most likely lower end of these ranges reflect a more correct flux and luminosity.

\begin{acknowledgements}

SB thanks an anonymous referee for very helpful comments that improved the manuscript.
The author acknowledges the $Chandra$ Observatory for performing the observations
of T Pyx and the Chandra Helpdesk for detailed answers to questions regarding particularly
the subpixel analysis.  SB acknowledges support
from T\"UB\.ITAK, The Scientific and Technological Research Council
of Turkey,  through project 108T735. SB also thanks S. Starfield, J.J. Drake,
M. Bode, J. Krautter, M. Hernanz, and J-U. Ness for support of the original
\cha\ T Pyx proposal.

\end{acknowledgements}

\end{document}